\documentclass[journal=jacsat,manuscript=article]{achemso}
\usepackage[T1]{fontenc}
\usepackage{verbatim}
\usepackage{chemformula}
\usepackage{amsmath}
\usepackage{amssymb}
\usepackage{siunitx}
\usepackage{hyperref}
\usepackage{cleveref}
\usepackage{multirow}
\usepackage{booktabs}
\usepackage{threeparttable}

\DeclareSIUnit{\au}{a.u.}

\newcommand{\edit}{ }

\title{Augmented Roothaan-Hall Hessian Applied to Spin-Restricted Open-Shell Density-Functional Theory}

\author{Yichi Zhang}
\affiliation{Department of Chemistry, The University of Hong Kong, Hong Kong 999077, P.R. China}
\alsoaffiliation{Laboratory for Synthetic Chemistry and Chemical Biology Limited, Hong Kong 999077, P.R. China}
\author{Jun Yang}
\email{juny@hku.hk}
\affiliation{Department of Chemistry, The University of Hong Kong, Hong Kong 999077, P.R. China}
\alsoaffiliation{Laboratory for Synthetic Chemistry and Chemical Biology Limited, Hong Kong 999077, P.R. China}
\alsoaffiliation
{State Key Laboratory of Synthetic Chemistry and HKU-CAS Joint Laboratory on New Materials, The University of Hong Kong, Hong Kong 999077, P.R. China}

\begin{document}

\maketitle

\begin{tocentry}
    \includegraphics[width=0.975\columnwidth]{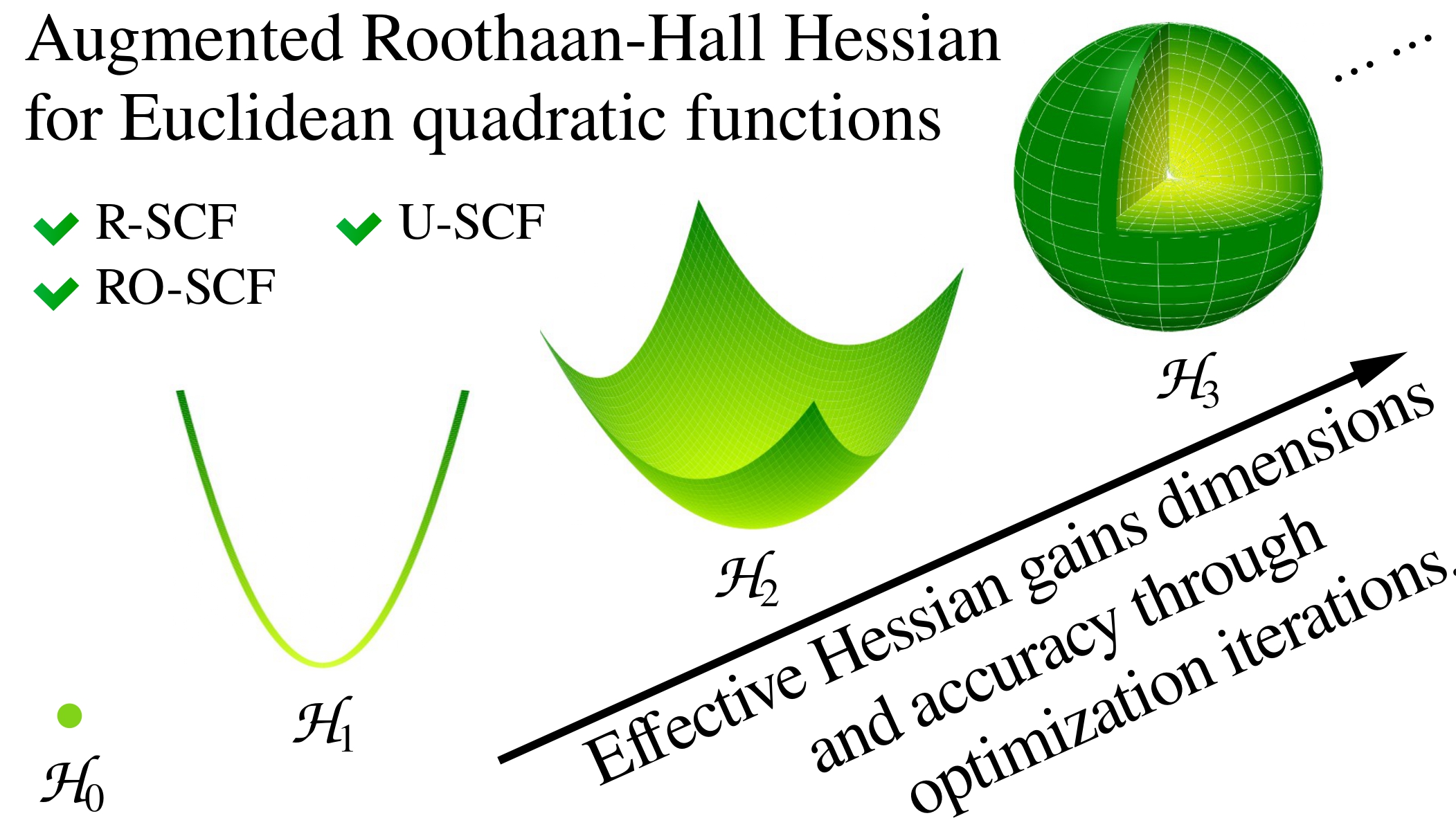}
\end{tocentry}

\begin{abstract}
We generalize the augmented Roothaan–Hall (ARH) Hessian formalism to the self-consistent field (SCF) optimization of spin-restricted open-shell (RO) wavefunctions, encompassing high-spin, low-spin, and two-determinant electronic states.
A detailed ARH formulation is presented.
We demonstrate that ARH is a highly efficient optimization algorithm for rapidly identifying accurate SCF solutions, primarily owing to its systematic construction of an effective Hessian, particularly in the case of Euclidean quadratic energy functions.
The ARH is built upon a universal energy formulation, including grid-based integration,
for spin-restricted closed-shell, spin-unrestricted and RO density functional theory (DFT), thereby unifying and simplifying their numerical implementation.
The performance of the present method is evaluated using two benchmarking studies.  
First, for a series of iron–sulfur clusters exhibiting different spin states, which represent notoriously challenging SCF problems, the ARH algorithm demonstrates superior convergence efficiency relative to L-BFGS and truncated Newton methods, requiring much fewer RO-SCF iterations to achieve convergence.  
Second, the ARH approach tends to avoids convergence to higher-energy stationary points in two-determinant RO-SCF calculations for singlet excited states of selected photoactive compounds.
Finally, an application of the ARH-based RO-SCF is illustrated by an investigation of the mechanistic origin of the spin-crossover phenomenon in Ni(II)-porphyrin complex utilized as a contrast agent.
\end{abstract}

\section{1 \ \ Introduction}

The spin-restricted open-shell (RO) approach provides a wavefunction framework for describing electronic states exhibiting an imbalance of $\alpha$ and $\beta$ electron spins that are not perfectly paired, as an alternative to spin-unrestricted (U) wavefunctions.
It yields a single set of spatial orbitals for both $\alpha$ and $\beta$ electrons,
making a reasonable representation of high-spin electronic structures\cite{rohf_original,simple}
and a reference state for higher-level quantum chemistry methods such as the mixed-reference spin-flip time-dependent density functional theory (DFT) \cite{mrsf1,mrsf2}.
Moreover, the DFT wavefunction for spin-pure singlet excited states upon single-electron excitation
requires a single spin-adapted RO configuration state function
that constitutes two Slater determinants for a pure singlet state ($S^2 = 0$).
To this end, we introduce the acronym 2D-DFT (two-determinant DFT) to the family of methods in this work and the literature\cite{wrong_xc,stokes,roks_decompose}, in contrast to standard single-determinant approaches.\edit

Despite its utility, RO self-consistent-field (SCF) wavefunctions are not in widespread use,
as they are mathematically complex to solve, leading to difficulties in  SCF convergence.
The traditional approach for solving RO-SCF involves diagonalizing the composite Fock matrix and updating orbitals by direct inversion in the iterative subspace (DIIS).\cite{rohf_original,pulay1980,pulay1982,mom,sgm}
A modern class of RO-SCF implementation leverages direct minimization on the flag manifold,\cite{high_spin1,flag1,flag2}\edit
where orbitals of separate spin components make up mutually orthogonal Fock subspaces.
The mainstream algorithms of this class include the limited-memory Broyden–Fletcher–Goldfarb–Shanno (L-BFGS)\cite{lbfgs1,lbfgs2,lbfgs3,low_spin}
and its variants for specific SCF problems\cite{soscf,gdm,gdm_rohf}.
Moreover, Newton's method that is formally quadratically convergent for high-spin RO-SCF has been implemented in \texttt{PySCF}.\cite{pyscf}
In this work, we aim to present a new scheme to solve RO-SCF, namely the augmented Roothaan-Hall (ARH) method,
which was first reported for spin-restricted (R) closed-shell SCF\cite{arh}
and later extended for unrestricted (U) and nuclear-electronic orbital SCF\cite{grassmann_scf}
as well as grand-canonical SCF\cite{gcscf}.
Here we discuss the theoretical basis of ARH in a more general picture than its original closed-shell framework.\cite{arh}
We demonstrate that ARH is more efficient than L-BFGS in the direct optimization of RO-SCF problems on the flag manifold,
as the objective functions of the electronic energy are quadratic or approximately quadratic in the Euclidean space.

The implementation of direct energy minimization requires an energy function and its orbital derivatives.
As there are four types of spin-constrained SCF states, including R, U and RO of high spin (ROH) and low spin (ROL),
it would be extremely tedious to implement their energy functions separately,
especially the evaluation of their grid-based integration of exchange-correlation functionals.\cite{dft1,dft2,dft3}
In this work, we develop a universal formulation to encompass all four spin types
from the tensorization viewpoint which adds the spin components as a new tensor dimension to all quantities,
such as density and Fock matrices.
The resulting compact and tidy expressions have enabled us to implement the RO-SCF energy function and orbital derivatives for both high spin and low spin states within a uniform implementation framework.
Moreover, the 2D-DFT energy function and orbital derivatives can be simply derived based on
the linear combination of those of the spin-contaminated singlet and triplet configurations.
As such, an RO spin-pure diradical state can also be solved for by direct energy minimization including ARH, which will be discussed in details.

The performance and applicability of our ARH-based RO-DFT formulation will be assessed on selected molecules for which traditional RO-DFT implementation is inefficient.
For instance, it is a well-known fact that the SCF orbital optimization of iron-sulfur clusters often encounters the convergence difficulties.
Here, our tests show that the ARH shortens the necessary iterations dramatically as compared with L-BFGS and truncated Newton's method.
Moreover, the ARH-based 2D-DFT carried out on selected photoactive compounds exhibits improved convergence behavior
and yields excitation energies comparable with the experimental data.
Lastly, we will showcase an application of the present method to Ni(II)-porphyrin complexes, a responsive contrasting agent for magnetic resonance imaging\cite{ni3}, where the coordination-induced spin crossover mechanism is correctly identified.
In view of those promising results, we recommend ARH as a superior solution to difficult RO-DFT problems.

\section{2 \ \ A General Spin Formulation for SCF Implementation}

In most electronic structure implementations, the procedures for conducting
spin-restricted closed-shell SCF,
spin-restricted open-shell SCF (of high spin and low spin)\edit and
spin-unrestricted SCF calculations
are formulated and coded independently,
including the corresponding two-electron integral contractions and exchange–correlation integrations.
In this section, we present a unified theoretical framework for the SCF implementation that consistently encompasses all four \edit spin formalisms.
Within this framework, spin-restricted open-shell DFT for both high-spin and low-spin states is treated as a special case of the general formulation.

The general SCF energy as a function of the occupied spatial molecular orbital coefficient matrix $\mathbf{C}$ is given by
\begin{equation}\begin{aligned}
    E(\mathbf{C}) = E^\text{HF}(\mathbf{C}) + E^\text{XC}(\mathbf{C}),
\end{aligned}\label{eq:energy}\end{equation}
where $E^\text{HF}(\mathbf{C})$ denotes the Hartree–Fock-like (HF) contribution and $E^\text{XC}(\mathbf{C})$ represents the exchange–correlation (XC) contribution.
The coefficient matrix, which depends on the specific wavefunction form, can be decomposed into at most three column partitions corresponding to the spin components listed in \cref{tab:spin}, i.e.,
\[
\mathbf{C} = \big[ \mathbf{C}^p \ \mathbf{C}^a \ \mathbf{C}^b \big],
\]
where the superscripts $p$, $a$ and $b$ stand for closed-shell, open-shell $\alpha$ and open-shell $\beta$ orbitals, respectively.\edit
This coefficient matrix is directly related to the occupation-number-free density matrix, which is defined as the outer product of the corresponding spin-block coefficient matrix,
\begin{equation}
    \mathbf{D}^w(\mathbf{C}^w)
    = \mathbf{C}^w \mathbf{C}^{w\top},\edit
\label{eq:density_matrix}\end{equation}
where the spin-component index $w$ takes values in $\{p, a, b\}$.  
Note that, in contrast to the convention in quantum chemistry,
the occupation numbers are separated from our definition of the density matrix for implementation convenience on flag manifolds.\edit
For convenience, one may also introduce a concatenated density matrix,
\[
\mathbf{D} = \big[ \mathbf{D}^p, \ \mathbf{D}^a, \ \mathbf{D}^b \big],
\]
which aggregates the contributions from all spin components.

\begin{table}\centering\begin{tabular}{l|cccc}
    \hline
    Spin types of wavefunctions & R & U & ROH & ROL \\
    \hline
    Spin components $w$ & $p$ & $a$, $b$ & $p$, $a$ & $p$, $a$, $b$ \\
    \hline
    Occupation numbers per orbital $n^w$ & 2 & 1, 1 & 2, 1& 2, 1, 1 \\
    \hline
\end{tabular}\caption{Spin components of four spin types of wavefunctions.}\label{tab:spin}\end{table}

Since $\mathbf{C}$ comprises orthogonal columns of one, two or three partitions
and the wavefunction is invariant to orbital rotations within each partition,
the SCF can be handled as a direct optimization problem on the flag manifold\cite{stiefel1,flag1,flag2}
(or the product of several flag manifolds).
The detailed optimization algorithms on the flag manifold are discussed in the supporting information of the work by \citet{gcscf}.
To make use of manifold optimization algorithms,
we need the Euclidean derivatives of \cref{eq:energy} with respect to $\mathbf{C}^w$.\cite{absil,boumal,grassmann}
It can be shown that the Euclidean gradient with respect to the density matrix is\cite{lehtola2020}
\begin{equation}\begin{aligned}
    \frac{\partial E}{\partial \mathbf{D}^w} = n^w \mathbf{F}^w
\end{aligned}\label{eq:d_grad}\end{equation}
where $n^w$ is the occupation number per orbital in the $w^\text{th}$ spin component.
(See \cref{tab:spin})
Moreover, the Euclidean gradient with respect to the coefficient matrix is thus\cite{lehtola2020}
\begin{equation}\begin{aligned}
    \frac{\partial E}{\partial \mathbf{C}^w}
    &= \left( \frac{\partial E}{\partial \mathbf{D}^w} + \frac{\partial E}{\partial \mathbf{D}^{w\top}} \right) \mathbf{C}^w \\
    &= 2 n^w \mathbf{F}^w \mathbf{C}^w
\end{aligned}\label{eq:c_grad}\end{equation}
where the Fock matrix $\mathbf{F}^w = \mathbf{F}^{\text{HF},w} + \mathbf{V}^{\text{XC},w}$ is the sum of the HF-like ($\mathbf{F}^{\text{HF},w}$) and XC ($\mathbf{V}^{\text{XC},w}$) contributions.
$\mathbf{F}^{\text{HF},w}$ and $\mathbf{V}^{\text{XC},w}$ are given in the following sections.
Further differentiating \cref{eq:d_grad} and \cref{eq:c_grad} leads to the Euclidean Hessian,
which is crucial for driving the second-order optimization,
\begin{equation}
    \sum_u \frac{\partial^2 E}{\partial \mathbf{C}^w \partial \mathbf{C}^u} : \dot{\mathbf{C}}^u
    = 2 \left[
        \left( \sum_u
            \frac{\partial^2 E}{\partial \mathbf{D}^w  \partial \mathbf{D}^u} : \dot{\mathbf{D}}^u
        \right) \mathbf{C}^w
        + n^w \mathbf{F}^w \dot{\mathbf{C}}^w
    \right],
\label{eq:c_hess}\end{equation}
where $\dot{\mathbf{C}}$ and $\dot{\mathbf{D}}$ are variations in the coefficient matrix and density matrix, respectively, 
which are related by differentiating \cref{eq:density_matrix} (also known as differential canonical projection)\cite{absil,boumal,grassmann},
\begin{equation}
    \dot{\mathbf{D}}^w
    = \mathbf{C}^w \dot{\mathbf{C}}^{w\top}
    + \dot{\mathbf{C}}^w \mathbf{C}^{w\top}.
\end{equation}
Here, the double dot contraction
$\frac{\partial^2 f}{\partial \mathbf{X} \partial \mathbf{Y}} :\mathbf{Z}$
denotes the matrix whose $(i,j)$ element is calculated as $\sum_{kl} \frac{\partial^2 f}{\partial X_{ij} \partial Y_{kl}} Z_{kl}$.
The Hessian with respect to the density matrix
is simply the first-order derivative of the Fock matrix, which is
also divided into the corresponding HF-like and XC components,
\begin{equation}
    \sum_u \frac{\partial^2 E}{\partial \mathbf{D}^w \partial \mathbf{D}^u} : \dot{\mathbf{D}}^u
    = n^w \dot{\mathbf{F}}^w(\dot{\mathbf{D}})
    = n^w \left[ \dot{\mathbf{F}}^{\text{HF},w}(\dot{\mathbf{D}}) + \dot{\mathbf{V}}^{\text{XC},w}(\dot{\mathbf{D}}) \right].
\label{eq:d_hess}\end{equation}

The preconditioner reported in \citet{low_spin} is utilized to obtain faster convergence of L-BFGS and the solution to the subproblem of truncated Newton's method and ARH.

In the following two sections we review the expressions of the HF-like and XC energies and Fock matrices.

\subsection{2.1 \ \ The HF-like part}

The HF-like energy is\cite{szabo,rohf_original,high_spin1,high_spin2,low_spin}
\begin{equation}
    E^\text{HF}(\mathbf{C}) := \frac{1}{2} \sum_w n^w \text{Tr}\left[
        (\mathbf{H}_\text{core} + \mathbf{F}^{\text{HF},w})
        \mathbf{D}^w
    \right],
\end{equation}
where $\mathbf{H}_\text{core}$ is the one-electron core Hamiltonian matrix
(the sum of the kinetic and nuclear-electron attraction matrices)
and $\mathbf{F}^{\text{HF},w}$ is the Fock matrix of the HF-like part,\cite{rohf_original,high_spin1,high_spin2,low_spin}
\begin{equation}\begin{aligned}
    \mathbf{F}^{\text{HF},p}(\mathbf{D})
    &:= \mathbf{H}_\text{core} + \mathbf{J}(\mathbf{D}) - c_\text{EXX} \left[
        \mathbf{K}(\mathbf{D}^p)
        + \frac{1}{2} \mathbf{K}(\mathbf{D}^a)
        + \frac{1}{2} \mathbf{K}(\mathbf{D}^b)
    \right] \\
    \mathbf{F}^{\text{HF},a}(\mathbf{D})
    &:= \mathbf{H}_\text{core} + \mathbf{J}(\mathbf{D}) - c_\text{EXX} \left[
        \mathbf{K}(\mathbf{D}^p)
        + \mathbf{K}(\mathbf{D}^a)
        + \lambda \mathbf{K}(\mathbf{D}^b)
    \right] \\
    \mathbf{F}^{\text{HF},b}(\mathbf{D})
    &:= \mathbf{H}_\text{core} + \mathbf{J}(\mathbf{D}) - c_\text{EXX} \left[
        \mathbf{K}(\mathbf{D}^p)
        + \lambda \mathbf{K}(\mathbf{D}^a)
        + \mathbf{K}(\mathbf{D}^b)
    \right],
\end{aligned}\label{eq:fock}\end{equation}
where $c_\text{EXX}$ is the percentage of the HF exact exchange.
The Coulomb and exchange matrices involved in \cref{eq:fock} are defined as
\begin{equation}\begin{aligned}
    J(\mathbf{D})_{\mu\nu} &:= \sum_{w\sigma\lambda} n^w D^w_{\sigma\lambda} (\mu\nu|\sigma\lambda) \\
    K(\mathbf{D}^w)_{\mu\nu} &:= \sum_{\sigma\lambda} D^w_{\sigma\lambda} (\mu\lambda|\nu\sigma).
\end{aligned}\end{equation}
An exchange matrix $\mathbf{K}(\mathbf{D}^w)$ is simply dropped from the expressions if $w$ does not belong to the chosen spin type.
The constant $\lambda$ is the spin coupling coefficient that regulates the exchange interaction between the unpaired $\alpha$ and $\beta$ electrons.
It takes the value of 0 for U-SCF and some other constant for RO-SCF decided by the specific spin-coupling scheme.\cite{coupling,orca_ro,low_spin}

Note that the equations in \cref{eq:fock} are linear functionals of the density matrix,
and thus the variations with respect to the density matrix are simply
\begin{equation}\begin{aligned}
    \dot{\mathbf{F}}^{\text{HF},p}(\dot{\mathbf{D}})
    &= \mathbf{J}(\dot{\mathbf{D}}) - c_\text{EXX} \left[
        \mathbf{K}(\dot{\mathbf{D}}^p)
        + \frac{1}{2} \mathbf{K}(\dot{\mathbf{D}}^a)
        + \frac{1}{2} \mathbf{K}(\dot{\mathbf{D}}^b)
    \right] \\
    \dot{\mathbf{F}}^{\text{HF},a}(\dot{\mathbf{D}})
    &= \mathbf{J}(\dot{\mathbf{D}}) - c_\text{EXX} \left[
        \mathbf{K}(\dot{\mathbf{D}}^p)
        + \mathbf{K}(\dot{\mathbf{D}}^a)
        + \lambda \mathbf{K}(\dot{\mathbf{D}}^b)
    \right] \\
    \dot{\mathbf{F}}^{\text{HF},b}(\dot{\mathbf{D}})
    &= \mathbf{J}(\dot{\mathbf{D}}) - c_\text{EXX} \left[
        \mathbf{K}(\dot{\mathbf{D}}^p)
        + \lambda \mathbf{K}(\dot{\mathbf{D}}^a)
        + \mathbf{K}(\dot{\mathbf{D}}^b)
    \right].
\end{aligned}\end{equation}

\subsection{2.2 \ \ The XC part}

The DFT grid-based integration over real space for all the spin components is necessary for XC energy evaluation and XC potential matrix formation.
When there are unpaired electrons,
the XC functional only recognizes the $\alpha/\beta$ scheme which focuses on the respective total densities of $\alpha$ and $\beta$,
instead of the $p/a/b$ scheme introduced in the HF part.
Therefore, one has to transform the $p/a/b$ density matrix to the $\alpha/\beta$ one,
\begin{equation}
    [\mathbf{D}^\alpha, \ \mathbf{D}^\beta]
    = [\mathbf{D}^p, \ \mathbf{D}^a, \ \mathbf{D}^b]
    \times \left[ \begin{array}{cc}
        1 & 1 \\
        1 & 0 \\
        0 & 1
    \end{array}
    \right],
\end{equation}
and the $\alpha/\beta$ potential matrix (obtained from the $\alpha/\beta$ density matrix) back to the $p/a/b$ one.
\begin{equation}
    [\mathbf{V}^{\text{XC},p}, \ \mathbf{V}^{\text{XC},a}, \ \mathbf{V}^{\text{XC},b}]
    = [\mathbf{V}^{\text{XC},\alpha}, \ \mathbf{V}^{\text{XC},\beta}]
    \times \left[ \begin{array}{ccc}
        \frac{1}{2} & 1 & 0 \\
        \frac{1}{2} & 0 & 1
    \end{array}
    \right].
\end{equation}
The variations in the density matrix and the potential matrix also follow the rules above.
Again, $\mathbf{D}^w$ and $\mathbf{V}^{\text{XC},w}$
(and their corresponding rows and columns of the transformation matrices)
are simply dropped from the expressions
if $w$ does not belong to the chosen spin type.

In the following, we discuss the evaluation of the $p/\alpha/\beta$ potential matrix constructed from the $p/\alpha/\beta$ density matrix.  
In this context, the spin-component indices $u$, $v$, and $w$ are taken among $p/\alpha/\beta$, rather than $p/a/b$.  
For the sake of notational brevity, we temporarily adopt the Einstein summation convention.

\begin{table}\centering\begin{tabular}{l|l}
    \hline
    Symbols & Definitions \\
    \hline
    $u$, $v$, $w$ & Spin component indices ($p$ for R and $\alpha/\beta$ for RO and U) \\
    $x$           & Spin component compound index ($pp$ for R and $\alpha\alpha/\alpha\beta/\beta\beta$ for RO and U)\edit \\
    $g$           & An index of a spatial grid \\
    $r$           & One of the three spatial dimensions \\
    $\mu$, $\nu$  & Indices of basis functions \\
    \hline
    \raisebox{-2pt}{$\dot{A}$} & The variation of the quantity $A$ with respect to the density matrix \\
    \hline
    $D^w_{\mu\nu}$             & The $(\mu,\nu)^\text{th}$ element of the density matrix of the spin type $w$ \\
    $E^\text{XC}$              & The total XC energy \\
    $V^{\text{XC},w}_{\mu\nu}$ & The $(\mu,\nu)^\text{th}$ element of the XC potential matrix of spin type $w$ \\
    $S^{wu}$                   & A matrix whose diagonal and off-diagonal elements are respectively 2 and 1 \\
    \hline
    $\phi^\mu_g$         & The $\mu^\text{th}$ basis value at grid $g$ \\
    $(\nabla\phi^w_g)_r$ & The $r^\text{th}$ spatial component of the gradient of $\phi^\mu_g$ \\
    $\rho^w_g$           & The density of the spin type $w$ at grid $g$ \\
    $(\nabla\rho^w_g)_r$ & The $r^\text{th}$ spatial component of the gradient of $\rho^w_g$ \\
    $\gamma^{wu}_g$      & The contracted density gradient of spin types $w$ and $u$ at grid $g$ \\
    $\epsilon_g$         & The exchange-correlation energy per electron at grid $g$ \\
    $f_g$                & The exchange-correlation energy density at grid $g$, $f_g = \epsilon_g \rho_g^w$ \\
    \hline
\end{tabular}\caption{Notation for grid integration for DFT.}\label{tab:symbol}\end{table}

The spatial density on grids can be computed as\cite{dft1,dft2,dft3}
\begin{equation}\begin{aligned}
    \rho^w_g &= n^w D^w_{\mu\nu} \phi^\mu_g \phi^\nu_g \\
    (\nabla\rho^w_g)_r &= 2 n^w D^w_{\mu\nu} (\nabla\phi^\mu_g)_r \phi^\nu_g \\
    \gamma^{wu}_g &= (\nabla\rho^w_g)_r (\nabla\rho^u_g)_r.
\end{aligned}\end{equation}
The occupation numbers are simply $n^p=2$ or $n^\alpha=n^\beta=1$.
With the spatial density, one can obtain the XC grid energy ($\epsilon_g$) and potentials ($\frac{\partial f_g}{\partial \rho^w_g}$ and $\frac{\partial f_g}{\partial \gamma^{wu}_g}$) with an XC library, such as \texttt{Libxc}\cite{libxc} and \texttt{XCFun}\cite{xcfun}.
The total XC energy and potential matrix can then be calculated by summing over all grid points as\cite{dft1,dft2,dft3,pyxdh}
\begin{equation}\begin{aligned}
    E^\text{XC}
    &= \epsilon_g \rho^w_g = f_g \\
    V^{\text{XC},w}_{\mu\nu}
    &= \frac{\partial f_g}{\partial \rho^w_g}
    \phi^\mu_g \phi^\nu_g
    + \left\{ S^{uv}
        \frac{\partial f_g}{\partial \gamma^{w u}_g}
        (\nabla\rho^w_g)_r \left[
            (\nabla\phi^\mu_g)_r \phi^\nu_g
            + (\nabla\phi^\nu_g)_r \phi^\mu_g
        \right]
    \right\}.
\end{aligned}\label{eq:xc1}\end{equation}
In \cref{eq:xc1} and others,
we assume that the grid weight\cite{lebedev} is already multiplied in the XC energy and potentials for notational conciseness.

For second-order algorithms that leverage the orbital Hessian,
one also needs to compute the variation of the spatial density and potential matrix with respect to the density matrix.
The variation of the spatial density is\cite{dft1,dft2,dft3}
\begin{equation}\begin{aligned}
    \dot{\rho}^w_g &= n^w \dot{D}^w_{\mu\nu} \phi^\mu_g \phi^\nu_g \\
    (\nabla\dot{\rho}^w_g)_r &= 2 n^w \dot{D}^w_{\mu\nu} (\nabla\phi^\mu_g)_r \phi^\nu_g \\
    \dot{\gamma}^{wu}_g &=
    (\nabla\dot{\rho}^w_g)_r (\nabla\rho^u_g)_r + 
    (\nabla\dot{\rho}^u_g)_r (\nabla\rho^w_g)_r ,
\end{aligned}\end{equation}
and that of the potential matrix is\cite{pyxdh}
\begin{equation}\begin{aligned}
    \dot{V}^{\text{XC},w}_{\mu\nu}
    &= \left(
        \frac{\partial^2 f_g}{\partial \rho^w_g \partial \rho^u_g} \dot{\rho}^u_g
        + \frac{\partial^2 f_g}{\partial \rho^w_g \partial \gamma^{uv}_g} \dot{\gamma}^{uv}_g
    \right) \phi^\mu_g \phi^\nu_g \\
    &+ S^{wu} \left[
        \left(
            \frac{\partial^2 f_g}{\partial \rho^v_g \partial \gamma^{wu}_g} \dot{\rho}^v_g
            + \frac{\partial^2 f_g}{\partial \gamma^x_g \partial \gamma^{wu}_g} \dot{\gamma}^x_g\edit
        \right) (\nabla\rho^u_g)_r
        + \frac{\partial f_g}{\partial \gamma^{wu}_g}
        (\nabla\dot{\rho}^u_g)_r
    \right] \left[
        (\nabla\phi^\mu_g)_r \phi^\nu_g
        + (\nabla\phi^\nu_g)_r \phi^\mu_g
    \right],
\end{aligned}\end{equation}
where the second-order XC potentials are also given by an XC library.
Note that only one spin component is written with $\gamma^x_g$ and $\dot{\gamma}^x_g$ in the term
$\frac{\partial^2 f_g}{\partial \gamma^x_g \partial \gamma^{wu}_g} \dot{\gamma}^x_g$,\edit
where the index $x$\edit is a compound one taking the value of $\alpha\alpha / \alpha\beta / \beta\beta$.
Given the tensorial grid-based integration,
instead of writing four separate implementations for four different spin types,
now we unify all spin cases in a single implementation once and for all.
We have also developed a code generator that automates the tensor arithmetic while making use of the tensor symmetry in XC evaluation.

\section{3 \ \ Two-determinant RO-DFT}

For singlet excited states
-- where one electron is promoted from a doubly occupied orbital to a virtual orbital without flipping its spin --
the one-determinant ROL wavefunction usually falls short.
The problem is fundamental: the $\alpha$ and $\beta$ electrons in the same spatial orbital are equally likely to be excited,
and a single determinant simply cannot capture this inherent indeterminacy.
In addition, a one-determinant ROL wavefunction suffers from strong spin contamination arising from the corresponding triplet state configuration.

As the pure singlet state ($S$) is a linear combination of the contaminated mixed state ($M$) and the triplet state ($T$) with $M_\text{S}=0$\cite{roks_decompose},\edit
the HF(-like) energy of the pure singlet excited state can be evaluated as
\cite{wrong_xc,stokes,roks_decompose}
\begin{equation}\begin{aligned}
    E^{\text{HF,}S}(\mathbf{C})
    &\equiv 2 E^{\text{HF,}M}(\mathbf{C}) - E^{\text{HF,}T}(\mathbf{C}),
\end{aligned}\label{eq:pure_s_hf_e}\end{equation}
where $\mathbf{C}$ is the single orbital set with two unpaired electrons.
By optimizing \cref{eq:pure_s_hf_e} rather than working solely with $E^{\text{HF,}M}(\mathbf{C})$,
and using the coefficient matrix $\mathbf{C}=[\mathbf{C}^p, \ \mathbf{C}^a, \ \mathbf{C}^b]$
(where $\mathbf{C}^a$ and $\mathbf{C}^b$ have only one column, respectively)
from the ROL wavefunction representation,
we directly obtain the HF energy of the pure singlet excited state.
In the evaluation of the contaminated singlet state energy $E^{\text{HF,}M}$,
the orbital $\mathbf{C}^a$ and $\mathbf{C}^b$ accommodate the unpaired $\alpha$ and $\beta$ electrons, respectively.
In the evaluation of the triplet energy $E^{\text{HF,}T}$, in contrast,
both $\mathbf{C}^a$ and $\mathbf{C}^b$ are treated as unpaired $\alpha$ orbitals.
The optimization problem is therefore defined on the flag manifold with three partitions,
which yields the 2D formalism for RO-HF optimization.

The 2D-DFT energy function analogue to \cref{eq:pure_s_hf_e} is\cite{wrong_xc}
\begin{equation}\begin{aligned}
    E^S_\text{Type I}
    &:= 2 E^M - E^T \\
    &= 2 E^{\text{HF,}M} - E^{\text{HF,}T}
    + 2 E^\text{XC}[\rho^{M,\alpha},\rho^{M,\beta}] - E^\text{XC}[\rho^{T,\alpha},\rho^{T,\beta}],
\end{aligned}\label{eq:type1}\end{equation}
which has been implemented in \texttt{Q-Chem} program package\cite{qchem4,qchem5}\edit and found practical applications\cite{stokes,charge_transfer,core,pcm_roks}.
The 2D-DFT can outperform the time-dependent (TD) DFT
since it takes into account the explicit orbital relaxation specific to that excited state,
which is altogether missing in TD-DFT.\cite{oodft,why}

On the other hand, in the XC terms of \cref{eq:type1},
neither $\rho^{M,\alpha/\beta}$ nor $\rho^{T,\alpha/\beta}$ reflects the real electron density of the pure singlet state.
The real electron density is supposed to be spin-unpolarized by equal contributions of the two determinants,
$| i\alpha, j\beta \rangle$ and $| j\alpha, i\beta \rangle$,
which can be calculated as
\begin{equation}
    \rho^S_g
    = \sum_{\mu\nu} \left( 2 D^p_{\mu\nu} + D^a_{\mu\nu} + D^b_{\mu\nu} \right) \phi^\mu_g \phi^\nu_g,\edit
\end{equation}
based on which we hereby propose another form of the XC energy,\edit
\begin{equation}
    E^S_\text{Type II} :=
    2 E^{\text{HF,}M} - E^{\text{HF,}T}
    + E^\text{XC}\left[\rho^S\right],\edit
\label{eq:type2}\end{equation}

We then coin the original \cref{eq:type1} as Type I (2D\textsubscript{I}-DFT) and the proposed \cref{eq:type2} as Type II (2D\textsubscript{II}-DFT).
There is currently no evidence about which type is more physical or accurate in treating the XC functional.

The density derivatives of 2D-DFT can be derived by applying \cref{eq:d_grad} and \cref{eq:d_hess} to both the contaminated singlet and triplet parts in \cref{eq:type1} and \cref{eq:type2}.
The orbital derivatives are then calculated with \cref{eq:c_grad} and \cref{eq:c_hess}.
The coupling coefficient $\lambda$ for 2D-DFT is 0, according to \citet{wrong_xc}.

\section{4 \ \ Augmented Roothaan-Hall Hessian}

In this section, we focus on the direct minimization algorithms of the RO-SCF energy discussed above.
In general, if the initial guess is fairly close to the solution,
Newton's method usually offers the most rapid convergence rate (in the sense of macro-iterations) among all direct optimization methods.
In addition to the gradient, Newton's method requires  Hessian information to construct a quadratic function for fitting the objective function in the proximity of the current iterate.
An update is subsequently determined by selecting the minimizer of the quadratic function.\cite{tr1,tr2,trah}

However, Newton's method has a severe bottleneck in the time-consuming Hessian evaluation.
As a remedy for this, the well-known L-BFGS algorithm uses an approximate inverse Riemannian Hessian iteratively built in the variable subspaces instead of the exact Hessian.\cite{lbfgs1,lbfgs2,lbfgs3}
During the optimization process,
the L-BFGS method  assembles an effective inverse Riemannian Hessian using previously computed gradient information along the iterate sequence,
and employs this approximation to determine the subsequent update.
L-BFGS and its variants have been reported for tackling the SCF problem of many kinds.\cite{soscf,gdm,gdm_rohf}
Despite the popularity of L-BFGS, the resulting effective inverse Riemannian Hessian does not necessarily gain accuracy by continuous update of iterates,
because the objective function is generally not a quadratic one on the Riemannian manifold and does not have a constant Hessian.

We have found that the augmented Roothaan-Hall (ARH) method is free of such a problem.\cite{arh}
ARH is especially useful in optimizing quadratic objective functions with constant Euclidean Hessian on matrix manifolds.
Rather than approximating the inverse Riemannian Hessian,
ARH adopts an indirect approach by building the effective Euclidean Hessian from the previous gradients and steps.
First of all, since the objective function is (approximately) quadratic in the Euclidean space,
the effective Euclidean Hessian becomes progressively accurate along the continuous update.
Secondly, although the Euclidean Hessian itself is approximate,
the further construction of the Riemannian Hessian from it,
using the orthogonal projection and the Weingarten map\cite{absil,weingarten,boumal},
is exact.
As a result, the approximation level is minimized and thus the effective Riemannian Hessian with the ARH method largely resembles the exact one.
Finally, the time complexity of the ARH-based Hessian costs just a small fraction of the exact one.\cite{gcscf}

The basic idea of ARH is as follows.
The Euclidean gradient $\mathcal{G}$ (2-dimensional) and Hessian $\mathcal{H}$ (4-dimensional) satisfy the gradient theorem (since $\mathcal{H}$ is the gradient of $\mathcal{G}$),
\begin{equation}
    \int_{\mathbf{A}}^{\mathbf{B}} \mathcal{H}(\mathbf{X}) : \text{d} \mathbf{X}
    \equiv \mathcal{G}(\mathbf{B}) - \mathcal{G}(\mathbf{A}),
\end{equation}
where $\mathbf{A}$ and $\mathbf{B}$ (both are 2-dimensional) are two positions in the Euclidean space.
For a quadratic function,
the Hessian is a constant that can be moved out of the integral,
\begin{equation}
    \mathcal{H} : (\mathbf{B} - \mathbf{A})
    = \mathcal{G}(\mathbf{B}) - \mathcal{G}(\mathbf{A}).
\label{eq:arh_hess1}\end{equation}
On the other hand, given a sequence of points $\left\{ \mathbf{X}_i\right\}$ and a reference point $\mathbf{X}$ in the Euclidean space,
we can approximate an arbitrary vector $\mathbf{\Delta}$ by projecting it to the subspace spanned by $\left\{ \bar{\mathbf{X}}_i\right\} := \left\{ \mathbf{X}_i - \mathbf{X} \right\}$,
\begin{equation}
    \Tilde{\mathbf{\Delta}} = \hat{P}(\mathbf{\Delta})
    := \sum_{ij} \bar{\mathbf{X}}_{i}
    \left(\mathbf{T}^{-1}\right)_{ij}
    \text{Tr}(\bar{\mathbf{X}}_{j}^\top\mathbf{\Delta}),
\label{eq:proj}\end{equation}
where the inner product is defined as the Frobenius product of matrices
\begin{equation}
    \langle \mathbf{\Phi} | \mathbf{\Psi} \rangle = \text{Tr}(\mathbf{\Phi}^\top \mathbf{\Psi})
\end{equation}
and $\mathbf{T}$ is the overlap matrix among the basis set $\left\{ \bar{\mathbf{X}}_i\right\}$
\begin{equation}
    T_{ij} := \text{Tr}(\bar{\mathbf{X}}_i^\top\bar{\mathbf{X}}_j),
\end{equation}
assuming that $\left\{ \bar{\mathbf{X}}_i\right\}$ is linearly-independent but not necessarily orthogonal.
Accordingly, the contraction between the Hessian $\mathcal{H}$ and the vector $\mathbf{\Delta}$ can be approximated by the following one between the Hessian and the projected vector $\tilde{\mathbf{\Delta}}$,
\begin{equation}
    \mathcal{H} : \mathbf{\Delta}
    \approx \mathcal{H} : \tilde{\mathbf{\Delta}}
    = \sum_{ij} \mathcal{H} : \bar{\mathbf{X}}_{i}
    \left(\mathbf{T}^{-1}\right)_{ij}
    \text{Tr}(\bar{\mathbf{X}}_{j}^\top\mathbf{\Delta}).
\label{eq:arh_hess2}\end{equation}
Inserting \cref{eq:arh_hess1} into \cref{eq:arh_hess2},
\begin{equation}
    \mathcal{H} : \mathbf{\Delta}
    \approx \mathcal{H} : \tilde{\mathbf{\Delta}}
    = \sum_{ij} \bar{\mathbf{G}}_{i}
    \left(\mathbf{T}^{-1}\right)_{ij}
    \text{Tr}(\bar{\mathbf{X}}_{j}^\top\mathbf{\Delta}),
\label{eq:arh_hess3}\end{equation}
where the gradient difference is
\begin{equation}
    \bar{\mathbf{G}}_i := \mathcal{G}(\mathbf{X}_i) - \mathcal{G}(\mathbf{X}).
\end{equation}
During the optimization, the current iterate is chosen as the reference $\mathbf{X}$, while the step vectors of the previous iterates form the sequence $\left\{ \bar{\mathbf{X}}_i\right\}$.
Their corresponding gradients can be retrieved from the memory so that $\left\{ \bar{\mathbf{G}}_i \right\}$ can be calculated.
In this way, the exact Hessian-vector product is replaced by the inexpensive sum over only a few element-wise products in \cref{eq:arh_hess3}.
In practice, only a few most recent $\left\{ \bar{\mathbf{X}}_i\right\}$ and $\left\{ \bar{\mathbf{G}}_i\right\}$ are stored, for lowering computing memory usage.

It is known that the HF energy is a quadratic function,
while the DFT energy can be viewed as an approximative one (because of the XC functional),
of the density matrix in the Euclidean space.
Therefore, the ARH Hessian perfectly fits the SCF problems.
To make use of the ARH Hessian,
we replace the density Hessian in \cref{eq:d_hess} with \cref{eq:arh_hess3} using the following variable substitutions:
\begin{itemize}
    \item \cref{eq:d_grad}, or its variant for 2D-DFT $\to \mathbf{G}$,
    \item \cref{eq:d_hess}, or its variant for 2D-DFT $\to \mathcal{H}$,
    \item $\bar{\mathbf{D}}_i = [\mathbf{D}^p_i-\mathbf{D}^p, \ \mathbf{D}^a_i-\mathbf{D}^a, \ \mathbf{D}^b_i-\mathbf{D}^b] \to \bar{\mathbf{X}}_i$.
    \item $\dot{\mathbf{D}} = [\dot{\mathbf{D}}^p, \ \dot{\mathbf{D}}^a, \ \dot{\mathbf{D}}^b] \to \mathbf{\Delta}$,
\end{itemize}
As the optimization proceeds,
the ARH Hessian spanned by $\left\{ \bar{\mathbf{D}}_i\right\}$ always (for HF) or usually (for DFT) becomes progressively accurate,
since the Hessian with respect to the density matrices is a constant (for HF) or nearly a constant (for DFT).

\section{5 \ \ Results and Discussions}

\begin{table}[h!]
    \centering
    \begin{tabular}{ccc}
        \hline
         & Macro-iterations & Micro-iterations \\
        \hline
        \multirow{2}{*}{L-BFGS} & Two-loop recursion      & Armijo backtracking line search \\
                                & (No Fock construction)  & (Fock construction)             \\
        \hline
        \multirow{3}{*}{Newton} & Calculating the current & Solving the \textbf{exact}-Hessian-gradient linear \\
                                & energy and gradient     & equation with truncated conjugate gradient         \\
                                & (Fock construction)     & (Fock construction)                                \\
        \hline
        \multirow{3}{*}{ARH}    & Calculating the current & Solving the \textbf{ARH}-Hessian-gradient linear \\
                                & energy and gradient     & equation with truncated conjugate gradient       \\
                                & (Fock construction)     & (No Fock construction)                           \\
        \hline
    \end{tabular}
    \caption{The functions of the macro- and micro-iterations of the three methods.}
    \label{tab:iter_purpose}
\end{table}

In this section, we compare the results based on our ARH implementation with those of other optimization schemes including L-BFGS and truncated Newton's method for two difficult cases --
RO-DFT calculation of iron(III)-sulfur clusters and
2D-DFT calculation of selected photoactive compounds.
As implemented in our \texttt{Chinium} package,
all the three methods invoke macro-iterations and micro-iterations in each macro-iteration.
The main tasks of these iterations are shown in \cref{tab:iter_purpose}.
Since the Fock-like matrix construction is most computationally intensive,
we compare the number of these iterative cycles where the Fock matrix formation is repeatedly invoked: the micro-iterations of L-BFGS,
the macro- and micro-iterations of truncated Newton's method and the macro-iterations of ARH.
The energy convergence threshold of the macro-iterations is maintained at \SI{e-10}{\au}.
For truncated Newton's method and ARH,
the threshold $\kappa$
(the ratio of the quadratic function increment to the lowered quadratic function)
of the micro-iterations is 0.001 and 0.01, respectively.
In each micro-iteration of L-BFGS,
the step size shrinks by 0.75,
until the Armijo's condition (with $c_1=0.1$) is fulfilled.
For L-BFGS and ARH, the Fock and density matrices of the last 20 macro-iterations are stored.
The SG-2 grid is used for grid-based integration.\cite{sg3}

\subsection{5.1 \ \ Multiple spin states of Iron(III)-sulfur clusters}

\begin{figure}
    \centering
    \includegraphics[width=\columnwidth]{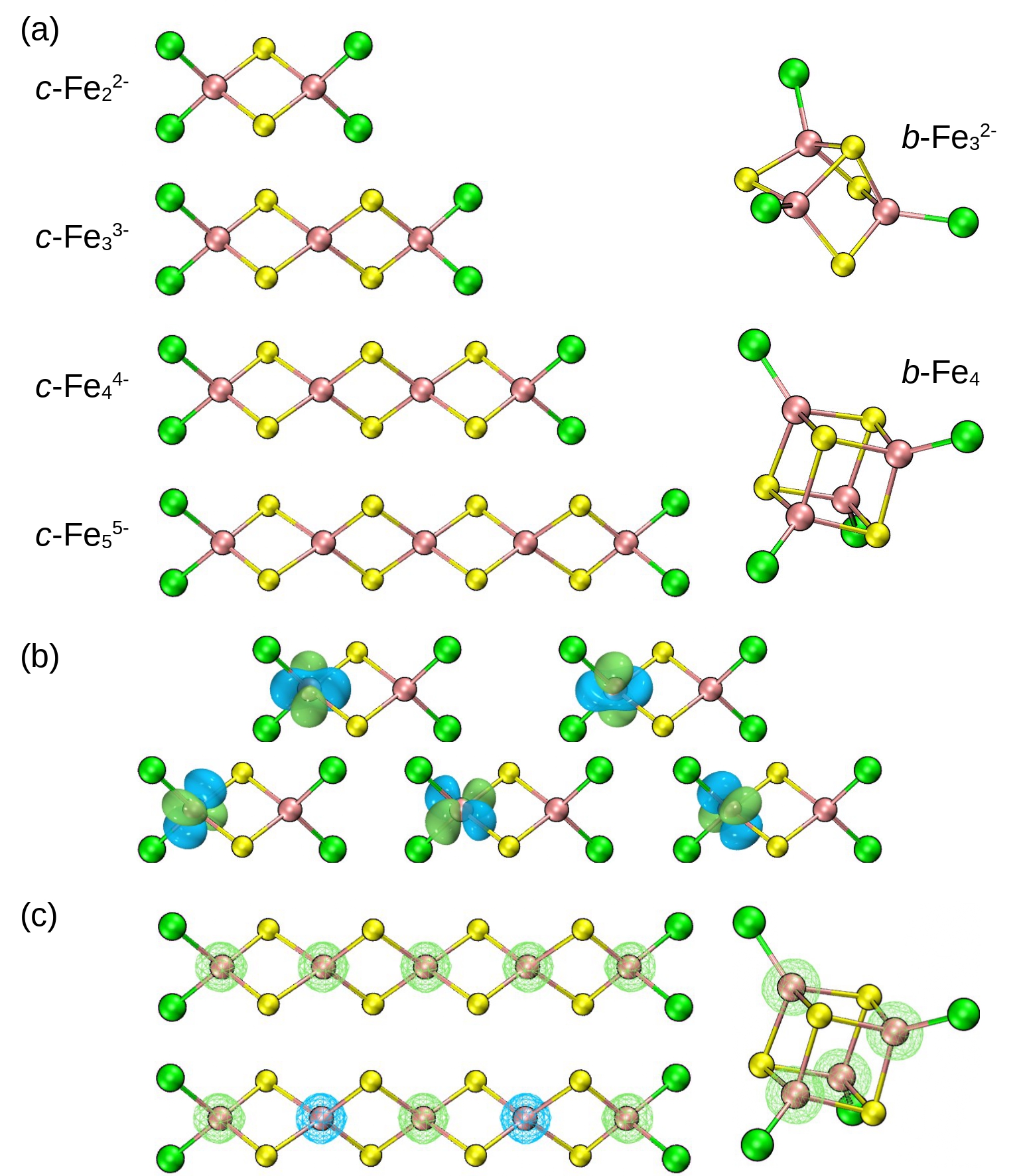}
    \caption{
        The tested iron-clusters (those with terminal \ch{-Cl} as examples)
        (a) The molecular models;
        (b) The initially guessed singly occupied orbitals,
        those of one \ch{Fe} atom of the \textit{c-}\ch{Fe2^{2-}} as examples.
        (c) The converged spin density, taking the \textit{c-}\ch{Fe5^{5-}} chain (top: ferromagnetic; bottom: anti-ferromagnetic) and the \textit{b-}\ch{Fe4} (ferromagnetic only) at the level of RO-B3LYP as examples.
        (Pink balls -- Fe; Yellow balls -- S; Green balls -- Cl;
        Isosurface: \SI{0.05}{\au};
        Green and blue surfaces -- two phases of orbitals;
        Green and blue wireframes -- two spin directions;
        Drawn with Multiwfn\cite{multiwfn1,multiwfn2} and VMD\cite{vmd})
    }
    \label{fig:fes}
\end{figure}

Our ARH optimization was applied to a variety of high- and low-spin states of the open-shell iron-sulfur clusters in \cref{fig:fes}a,
including four chain-like (\textit{c}-) clusters and two bulky (\textit{b}-) ones,
with two types of terminal groups, \ch{-Cl} or \ch{-SH}.
These open-shell systems are challenging to standard DFT calculations which are susceptible to complete convergence failures or false convergence to high-energy states due to the dense spin states.

The geometries of these clusters were optimized at the level of high-spin U-B3LYP\cite{b3lyp}/6-31G(d)\cite{6-31g*} using the \texttt{Gaussian}\cite{g16} program.
Next, RO-HF and RO-B3LYP single-point calculations were carried out on these clusters with the same basis set.
The initial guess of the molecular orbitals for the following RO-SCF was obtained by localizing the SAP (superposition of atomic potentials)\cite{sap} orbitals and identifying the singly occupied \ch{Fe}(III) $d$ orbitals, as shown in \cref{fig:fes}b.
Therefore, all the five $d$ orbitals of an \ch{Fe}(III) ion are assumed to host parallel spins initially,
with each \ch{Fe}(III) ion having either $M_S=\frac{5}{2}$ (symbolized by $\uparrow$) or $M_S=-\frac{5}{2}$ (symbolized by $\downarrow$).
The total spin states were then initialized according to the spins on all Fe sites,
for example, $\uparrow\downarrow\downarrow\downarrow\uparrow$
for each of the two terminal \ch{Fe} sites with $M_S=\frac{5}{2}$ and each of the three middle \ch{Fe} sites with $M_S=-\frac{5}{2}$.
The converged spin densities, as illustrated in \cref{fig:fes}c, are found aligned with the initial spin states.
Finally, we carried out complete-active-space (CAS) SCF/6-31G(d) calculations on \textit{c}-\ch{Fe2^{2-}} and \textit{c}-\ch{Fe3^{3-}},
with the $d$ electrons / orbitals of the \ch{Fe}(III) ion
(i.e., (10e, 10o) for \textit{c}-\ch{Fe2^{2-}}
and (15e, 15o) for \textit{c}-\ch{Fe3^{3-}})
as the active spaces,
for a reasonable comparison between the high-spin and low-spin energies,
using the \texttt{ORCA}\cite{orca1,orca2} program.\edit

The resulting numbers of iterations spent by the three methods are listed in \cref{tab:benchmark}.
It can be seen that L-BFGS convergence requires up to hundreds of iterations, exhibiting the largest iteration numbers and thus the lowest efficiency.
Truncated Newton's method shortens the iteration to roughly $\frac{1}{5}\sim\frac{1}{3}$ of that of L-BFGS,
and ARH reduces the iteration numbers to roughly $\frac{1}{3}$ of that of the Newton's method.
Furthermore, the robustness of ARH is evidenced by its nearly constant iteration counts across different spin types.
More detailed analysis of the ARH performance for RO-HF and RO-B3LYP indicates that
ARH is more computationally efficient in conjunction with RO-HF than with RO-B3LYP.
The performance difference between RO-HF and RO-DFT originates from the fact that, in the latter case, the objective function is no longer strictly quadratic due to the contribution of the XC functional.

\begin{table}[h!]
    \caption{
        Comparison of the iteration numbers between various RO-SCF optimizations on the \ch{Fe-S} clusters.
        For each spin type and SCF algorithm, both the iteration numbers are given for \ch{-Cl} and \ch{-SH} terminal groups, respectively.
    }
    \centering
    \begin{threeparttable}\begin{tabular}{lccccccc}
        \hline
        \multirow{2}{*}{Cluster} &
        \multirow{2}{*}{Spin type} &
        \multicolumn{3}{c}{RO-HF} &
        \multicolumn{3}{c}{RO-B3LYP}  \\
        \cmidrule(r){3-5} \cmidrule(r){6-8}
        &&L-BFGS\tnote{a} & Newton\tnote{b} & ARH\tnote{c} &
        L-BFGS\tnote{a} & Newton\tnote{b} & ARH\tnote{c}  \\ \hline
        \multirow{2}{*}{\textit{c}-\ch{Fe2^{2-}}} & $\uparrow\uparrow$ & 135, 265 & 47, 80 & 15, 16 & 127, 108 & 71, 77 & 23, 28 \\
                                        & $\uparrow\downarrow$ & 154, 132 & 47, 66 & 14, 16 & 167, 133 & 72, 92 & 26, 28 \\
        \hline
        \multirow{2}{*}{\textit{c}-\ch{Fe3^{3-}}} & $\uparrow\uparrow\uparrow$ & 259, >330\tnote{d} & 56, 62 & 16, 16 & >331\tnote{d}, >355\tnote{d} & 84, 88 & 29, 27 \\
                                        & $\uparrow\downarrow\uparrow$ & 281, 279 & 55, 62 & 15, 16 & >337\tnote{d}, >331\tnote{d} & 73, 89 & 27, 25 \\
        \hline
        \multirow{4}{*}{\textit{c}-\ch{Fe4^{4-}}} & $\uparrow\uparrow\uparrow\uparrow$ & 229, >363\tnote{d} & 73, 84 & 21, 23 & >401\tnote{d}, >731\tnote{d} & 98, 114 & 37, 38 \\
                                        & $\uparrow\uparrow\downarrow\downarrow$ & 183, 300 & 74, 99 & 21, 22 & >394\tnote{d}, >463\tnote{d} & 102, 107 & 40, 37 \\
                                        & $\uparrow\downarrow\downarrow\uparrow$ & 134, >403\tnote{d} & 95, 76 & 21, 21 & >372\tnote{d}, >458\tnote{d} & 106, 116 & 41, 41 \\
                                        & $\uparrow\downarrow\uparrow\downarrow$ & 168, >436\tnote{d} & 88, 77 & 21, 21 & >605\tnote{d}, >478\tnote{d} & 111, 103 & 42, 39 \\
        \hline
        \multirow{4}{*}{\textit{c}-\ch{Fe5^{5-}}} & $\uparrow\uparrow\uparrow\uparrow\uparrow$ & >505\tnote{d}, 273 & 84, 98 & 23, 21 & >351\tnote{d}, >452\tnote{d} & 108, 95 & 42, 40 \\
                                        & $\uparrow\downarrow\downarrow\downarrow\uparrow$ & >648\tnote{d}, 136 & 112, 94 & 25, 20 & >474\tnote{d}, 413 & 132, 95 & 48, 36 \\
                                        & $\uparrow\uparrow\downarrow\uparrow\uparrow$ & >693\tnote{d}, 248 & 110, 78 & 25, 22 & >577\tnote{d}, 336 & 134, 110 & 42, 32 \\
                                        & $\uparrow\downarrow\uparrow\downarrow\uparrow$ & >396\tnote{d}, 233 & 85, 89 & 23, 22 & >597\tnote{d}, 433 & 136, 97 & 41, 36 \\
        \hline
        \textit{b}-\ch{Fe3^{2-}} & $\uparrow\uparrow\uparrow$ & >398\tnote{d}, >390\tnote{d} & 67, 283\tnote{e} & 19, 22 & 322, >606\tnote{d} & 82, 100 & 32, 37 \\
        \hline
        \textit{b}-\ch{Fe4} & $\uparrow\uparrow\uparrow\uparrow$ & 332, 285 & 73, 79 & 15, 18 & 380, 233 & 78, 88 & 28, 37 \\
        \hline
    \end{tabular}
    \begin{tablenotes}
        \item[a] Micro-iterations are counted.
        \item[b] Both macro- and micro-iterations are counted.
        \item[c] ARH for restricted open-shell wavefunctions is the algorithm newly proposed in this work. Macro-iterations are counted.
        \item[d] Not converged within 300 macro-iterations.
        \item[e] Convergence to a higher energy state.
    \end{tablenotes}
    \end{threeparttable}
    \label{tab:benchmark}
\end{table}

Although ARH exhibits considerably improved optimization performance when applied to RO calculations, it is important to emphasize that the use of ROL wavefunctions themselves must be carefully scrutinized in practice.
ROL states typically signal pronounced multi-reference characters with strong static correlations between the unpaired $\alpha$ and $\beta$ electrons.
Consequently, single-determinant approaches such as ROL-DFT are insufficient to stabilize ROL states and systematically overestimate the energies of low-spin states.
For instance, using multi-reference methods, iron–sulfur clusters have been demonstrated to possess low-spin, antiferromagnetically coupled ground states both in literature\cite{fes,orca_ro,low_spin} and our CAS-SCF calculations in \cref{tab:energy}.\edit
In contrast, as shown in \cref{tab:energy}, electronic configurations with greater high-spin character are predicted to be lower in energy at the RO-HF and RO-B3LYP levels of theory.  
On this basis, we strongly advise that ROL wavefunctions be employed with appropriate caution and subject to thorough validation in studies of systems with significant static correlations.

\begin{table}[h!]
    \caption{
        The low-spin electronic energies (\si{eV}) of the chain-like \ch{Fe-S} clusters relative to the high-spin ones.\edit
        For each spin type and SCF algorithm, both the energies are given for \ch{-Cl} and \ch{-SH} terminal groups, respectively.
    }
    \centering
    \begin{threeparttable}\begin{tabular}{lcccc}
        \hline
        Cluster & Spin type & RO-HF & RO-B3LYP & CAS-SCF\tnote{a} \\ \hline
        \textit{c}-\ch{Fe2^{2-}} & $\uparrow\downarrow$ & 0.080, 0.075 & 0.248, 0.224 & -0.060, -0.073 \\
        \hline
        \textit{c}-\ch{Fe3^{3-}} & $\uparrow\downarrow\uparrow$ & 0.126, 0.122 & 0.401, 0.385 & -0.075, -0.084 \\
        \hline
        \multirow{3}{*}{\textit{c}-\ch{Fe4^{4-}}} & $\uparrow\uparrow\downarrow\downarrow$ & 0.059, 0.059 & 0.209, 0.208 & -- \\
                                                                                    & $\uparrow\downarrow\downarrow\uparrow$ & 0.114, 0.110 & 0.360, 0.346 & -- \\
                                                                                   & $\uparrow\downarrow\uparrow\downarrow$ & 0.169, 0.165 & 0.535, 0.522 & -- \\
        \hline
        \multirow{3}{*}{\textit{c}-\ch{Fe5^{5-}}} & $\uparrow\downarrow\downarrow\downarrow\uparrow$ & 0.103, 0.100 & 0.314, 0.303 & -- \\
                                                                                        & $\uparrow\uparrow\downarrow\uparrow\uparrow$     & 0.102, 0.102 & 0.358, 0.357 & -- \\
                                                                                & $\uparrow\downarrow\uparrow\downarrow\uparrow$   & 0.201, 0.197 & 0.643, 0.633 & -- \\
        \hline
    \end{tabular}
    \begin{tablenotes}
        \item[a] Only the total spin multiplicities (instead of the spin carried by each \ch{Fe}(III) iron) are constrained in obtaining the CAS-SCF wavefunctions.
    \end{tablenotes}
    \end{threeparttable}
    \label{tab:energy}
\end{table}

\subsection{5.2 \ \ Singlet excited states of photoactive compounds}

\begin{figure}
    \centering
    \includegraphics[width=0.5\linewidth]{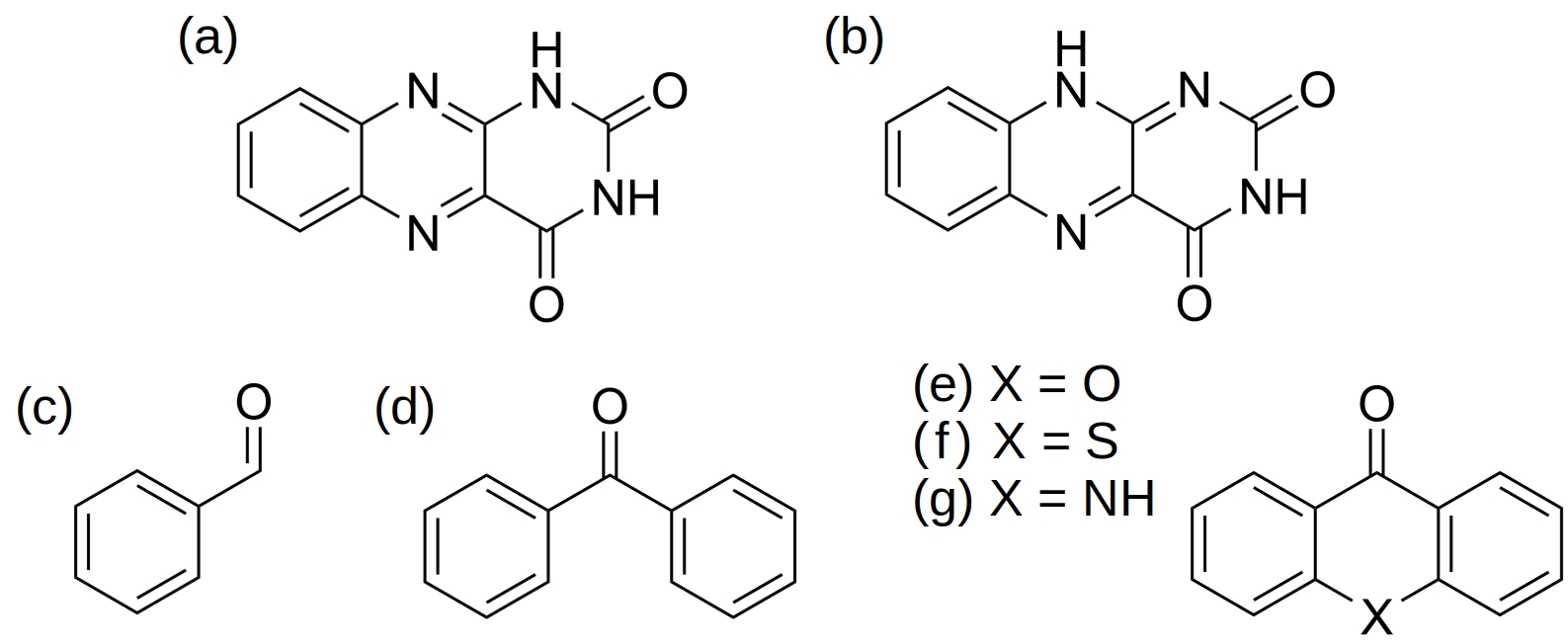}
    \caption{
        photoactive compounds:
        (a) Alloxazine ($C_\text{s}$);
        (b) Isoalloxazine ($C_\text{s}$);
        (c) Benzaldehyde ($C_\text{s}$);
        (d) Benzophenone ($C_2$);
        (e) Xanthone ($C_\text{2v}$);
        (f) Thioxanthone ($C_\text{2v}$);
        (g) Acridone ($C_\text{2v}$).
    }
    \label{fig:photo}
\end{figure}

We further evaluate the 2D-DFT performance in determining excited singlet states of organic photoactive molecules.
The testing molecules with bright one-electron excitation dominated by HOMO-to-LUMO transition are selected and shown in \cref{fig:photo}.
The experimental UV/Vis spectra of these molecules are available for comparison with our 2D-DFT excitation energies.
Their geometries were optimized at the level of B3LYP/cc-pVTZ\cite{cc-pvtz},
followed by single-point calculation at the level of 
2D\textsubscript{I}-B3LYP, 2D\textsubscript{II}-B3LYP, TD-B3LYP\cite{td1,td2,td3} and EOM-CCSD\cite{eom1,eom2},
using the same basis set.
The geometry optimization, TD-DFT and EOM-CCSD calculations were carried out using \texttt{Gaussian}.
Our \texttt{Chinium} package was used to conduct 2D\textsubscript{I}- and 2D\textsubscript{II}-B3LYP with L-BFGS, truncated Newton's method and ARH,
while \texttt{Q-Chem}\cite{qchem4,qchem5}\edit was used to carry out 2D\textsubscript{I}-B3LYP using the maximal-overlap method (MOM)\cite{mom} and squared-gradient minimization (SGM)\cite{sgm}.
For 2D-B3LYP, two sets of initial orbitals were tested:
the SAP orbitals (available in \texttt{Chinium}) and
the converged R-B3LYP orbitals.
The two frontier orbitals are both singly occupied.

\begin{table}
    \caption{Comparison of the iteration numbers between different optimization algorithms of 2D\textsubscript{I}-B3LYP for the photoactive compounds.}
    \centering
    \begin{threeparttable}\begin{tabular}{lcccccccc}
        \hline
        \multirow{2}{*}{\ \ \ Molecule} &
        \multicolumn{3}{c}{SAP guess} &
        \multicolumn{5}{c}{R-B3LYP guess}  \\
        \cmidrule(r){2-4} \cmidrule(r){5-9}
        &L-BFGS\tnote{a} & Newton\tnote{b} & ARH\tnote{c} &
        L-BFGS\tnote{a} & Newton\tnote{b} & ARH\tnote{c} & MOM & SGM \\ \hline
        a. Alloxazine           & >359\tnote{d} & 608 & 166 & 37 & 111 & 29 & 44 & >300\tnote{d} \\
        b. Isoalloxazine      & >353\tnote{d} & 207 & 84 & 50 & 121 & 39 & 31 & >300\tnote{d} \\
        c. Benzaldehyde    & 165 & 181 & 53 & 25 & 53 & 15 & 19 & >300\tnote{d} \\
        d. Benzophenone  & 203 & 187 & 46 & 37 & 122 & 28 & 28 & >300\tnote{d} \\
        e. Xanthone             & 36\tnote{e} & 120\tnote{e} & 41\tnote{e} & 32 & 87 & 27 & 30 & >300\tnote{d} \\
        f. Thioxanthone    & >445\tnote{d} & 166 & 48 & 27 & 81 & 25 & 26 & >300\tnote{d} \\
        g. Acridone               & 40\tnote{e} & 232 & 100 & 34 & 89 & 29 & 21 & >300\tnote{d} \\
        \hline
    \end{tabular}
    \begin{tablenotes}
        \item[a] Micro-iterations are counted.
        \item[b] Both macro- and micro-iterations are counted.
        \item[c] Macro-iterations are counted.
        \item[d] Not converged within 300 iterations.
        \item[e] Converged to a higher energy state.
    \end{tablenotes}
    \end{threeparttable}
    \label{tab:2d1}
\end{table}

\begin{table}
    \caption{Comparison of the iteration numbers between different optimization algorithms of 2D\textsubscript{II}-B3LYP for the photoactive compounds.}
    \centering
    \begin{threeparttable}\begin{tabular}{lcccccc}
        \hline
        \multirow{2}{*}{\ \ \ Molecule} &
        \multicolumn{3}{c}{SAP guess} &
        \multicolumn{3}{c}{R-B3LYP guess}  \\
        \cmidrule(r){2-4} \cmidrule(r){5-7}
        &L-BFGS\tnote{a} & Newton\tnote{b} & ARH\tnote{c} &
        L-BFGS\tnote{a} & Newton\tnote{b} & ARH\tnote{c}  \\ \hline
        a. Alloxazine           & 261 & 208 & 112 & 37 & 79 & 28 \\
        b. Isoalloxazine      & >442\tnote{d} & 251 & 115 & 104 & 136 & 51 \\
        c. Benzaldehyde    & 132 & 200 & 60 & 23 & 43 & 15 \\
        d. Benzophenone  & 138 & 167 & 62 & 35 & 103 & 25 \\
        e. Xanthone             & 38\tnote{e} & 257 & 104 & 30 & 82 & 29 \\
        f. Thioxanthone    & >395\tnote{d} & 165 & 61 & 28 & 76 & 34 \\
        g. Acridone               & 40\tnote{e} & 168 & 69 & 29 & 80 & 29 \\
        \hline
    \end{tabular}
    \begin{tablenotes}
        \item[a] Micro-iterations are counted.
        \item[b] Both macro- and micro-iterations are counted.
        \item[c] Macro-iterations are counted.
        \item[d] Convergence not reached within 300 macro-iterations.
        \item[e] Converged to a higher energy state.
    \end{tablenotes}
    \end{threeparttable}
    \label{tab:2d2}
\end{table}

\begin{table}
    \caption{Singlet excitation energies (eV) and oscillator strengths (\si{\au}) (in the parentheses if available) of the photoactive compounds.}
    \centering
    \begin{threeparttable}\setlength{\tabcolsep}{5pt}{\begin{tabular}{lccccccc}
        \hline
        \ \ \ Molecule             & Irrep     & 2D\textsubscript{I}-B3LYP & 2D\textsubscript{II}-B3LYP & TD-B3LYP       & EOM-CCSD      & Exp. \\
        \hline
        a. Alloxazine           & $A'$     & \phantom{-}3.02 & 3.16                 & 3.49 (0.0652) & 3.91 (0.1556) & 3.33\cite{alloxazine} \\
        b. Isoalloxazine     & $A'$     & \phantom{-}2.62 & 2.74                 & 3.12 (0.1675) & 3.54 (0.2784) & 2.83\cite{isoalloxazine}\tnote{a}\\
        c. Benzaldehyde   & $A''$   & \phantom{-}3.63 & 4.00                 & 3.68 (0.0001) & 4.02 (0.0002) & 3.78\cite{benzaldehyde}\edit \\
        d. Benzophenone & $A$      & \phantom{-}3.48 & 3.77                 & 3.61 (0.0009) & 3.95 (0.0010) & 3.73\cite{benzophenone} \\
        e. Xanthone            & $A_1$ & \phantom{-}3.46 & 3.58                 & 3.92 (0.0612) & 4.40 (0.0935) & 3.65\cite{xanthone} \\
        f. Thioxanthone   & $A_1$  & \phantom{-}3.13 & 3.26                 & 3.50 (0.0560)  & 4.04 (0.0868) & 3.26\cite{thioxanthone} \\
        g. Acridone             & $A_1$  & \phantom{-}3.07 & 3.18                 & 3.57 (0.0681)  & 3.97 (0.1118) & 3.10\cite{acridone} \\
        \hline
         \multicolumn{2}{c}{Mean error} & -0.18 & 0.00         & 0.17\phantom{ (0.0681)}                   & 0.59\phantom{ (0.0681)}                   & -- \\
         \multicolumn{2}{c}{Mean absolute error} & \phantom{-}0.18 & 0.10               & 0.24\phantom{ (0.0681)}                   & 0.59\phantom{ (0.0681)}                   & -- \\
        \hline
    \end{tabular}}
    \begin{tablenotes}
        \item[a] The experimental datum is of 3,10-dimethyl isoalloxazine.
    \end{tablenotes}
    \end{threeparttable}
    \label{tab:excitation_energy}
\end{table}

The iteration counts are reported in \cref{tab:2d1} and \cref{tab:2d2}, and display a pronounced sensitivity to the choice of the initial orbital guess.
The use of R-B3LYP orbitals as the starting point substantially reduces the number of iterations relative to the crude SAP guess.

Among the three direct minimization algorithms examined in this work, 
L-BFGS exhibits the least favorable performance,
characterized by a comparatively large number of iterations
and a high chance of converging to higher-lying stationary states.
The truncated Newton method yields iteration counts similar to those of L-BFGS,
but is less susceptible to the non-convex nature of the energy functional
and to convergence toward higher-order critical points.
In contrast, ARH consistently outperforms both L-BFGS and truncated Newton,
requiring the fewest iterations and displaying a strong propensity to converge to local minima.
For calculations initiated from R-B3LYP orbitals,
its performance is also comparable to that of MOM.
By comparison, SGM is the least efficient scheme among all methods examined,
as it reaches the maximum iteration limit even when R-B3LYP orbitals are employed as the initial guess.

The computed excitation energies in \cref{tab:excitation_energy}
are increasingly accurate as compared to experimental data when evaluated using EOM-CCSD, TD-B3LYP, 2D\textsubscript{I}-B3LYP and 2D\textsubscript{II}-B3LYP methods.
EOM-CCSD significantly overestimates the excitation energies,
indicating possible errors arising from the high-order correlations beyond doubles.
The error of TD-B3LYP is of average accuracy for excitation energy.
Better results are obtained by 2D\textsubscript{I}-B3LYP and 2D\textsubscript{II}-B3LYP methods.
Except for benzaldehyde,
the 2D\textsubscript{II}-B3LYP results are closer to the experimental data
than the 2D\textsubscript{I}-B3LYP ones,
which suggests that 2D\textsubscript{II}-DFT energy formalism appears to systematically address the underestimation of 2D\textsubscript{I}-DFT excitation energies.\edit

\subsection{5.3 \ \ Coordination-induced spin crossover of Ni(II)-porphyrin}

\begin{figure}
    \centering
    \includegraphics[width=\columnwidth]{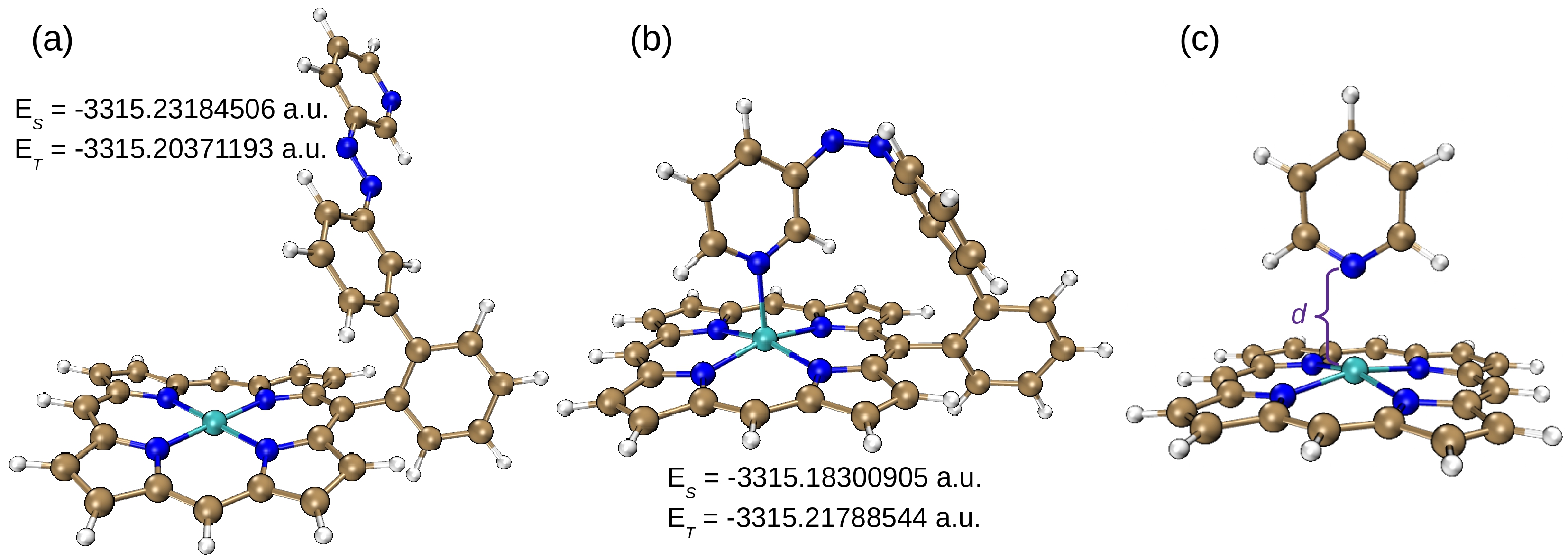}
    \caption{
        The azopyridine-substituted Ni-porphyrin (APSNP) with --\ch{C6H5} groups is replaced by \ch{H} atom for structure simplification:
        (a) \textit{trans}-APSNP with low spin;
        (b) \textit{cis}-APSNP with high spin;
        (c) the simplified model.
        (Cyan balls -- Ni; Brown balls -- C; Blue balls -- N; White balls -- H)
        (Drawn with VMD)
    }
    \label{fig:ni}
\end{figure}

Contrasting agents are crucial in magnetic resonance imaging.
The highly paramagnetic Gd agent is a traditional contrast agent.
In 2015, based on their previous work\cite{ni1,ni2}, \citet{ni3} reported the azopyridine-substituted Ni-porphyrin (APSNP) as a contrasting agent.
It features spin crossover induced by the change in the coordination number of the Ni(II) center stemming from the photo-isomerization of a pyridinylazophenyl structure.
In the \textit{trans} conformation (\cref{fig:ni}a), the pyridinyl substituent remains spatially distant from the Ni(II) center, which adopts a square-planar coordination geometry with the porphyrin ligand and exhibits a singlet ground state.
In the \textit{cis} conformation (\cref{fig:ni}b), the pyridinyl substituent coordinates to Ni(II), thereby inducing a square-pyramidal coordination geometry and stabilizing a triplet spin state.
The inter-conversion between \textit{trans} and \textit{cis} geometries is activated when the light of different wavelengths is absorbed.
The system acts as a photoswitch: 500 nm light drives a transition to the paramagnetic \textit{cis} state (``on''), whereas 435 nm light reverses the process to yield the diamagnetic \textit{trans} state (``off'').

\begin{figure}
    \centering
    \includegraphics[width=\columnwidth]{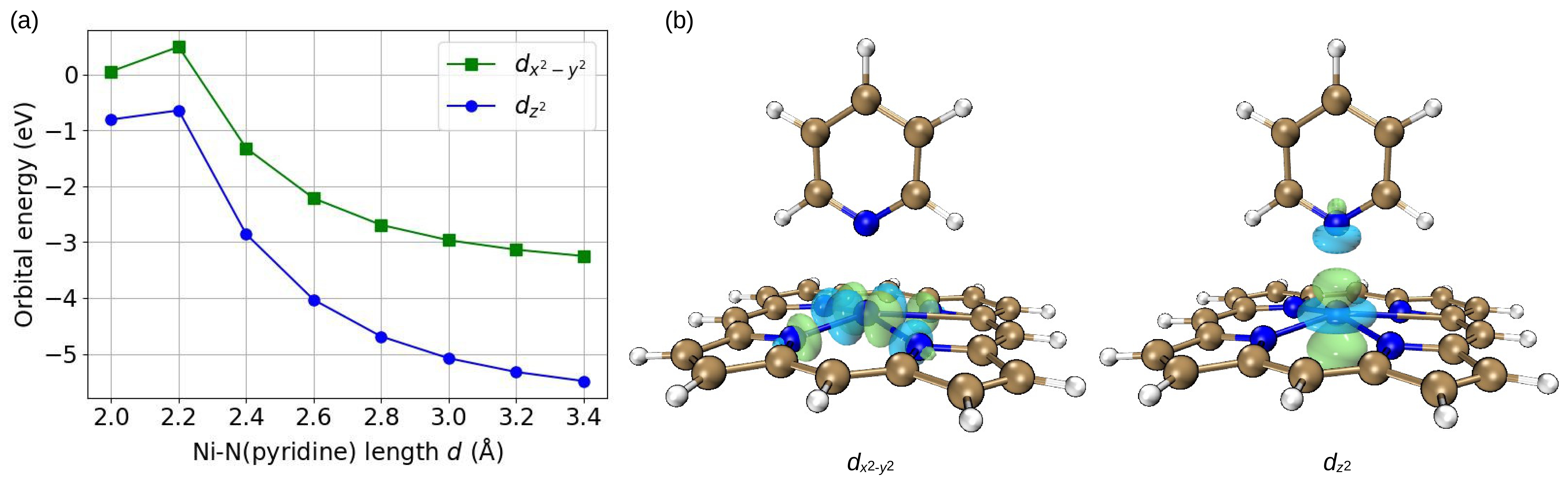}
    \caption{
        (a) The energy dependency of Ni(II) $d_{x^2-y^2}$ and $d_{z^2}$ orbitals on the highlighted \ch{Ni-N} bond length in the simplified model.
        (b) The two orbitals at $d=2.2$\AA. (Drawn with Multiwfn and VMD)
    }
    \label{fig:orbital}
\end{figure}

To illustrate a ``toy'' application of our new algorithm,
we apply RO-SCF to APSNP and its simplified model (\cref{fig:ni}c)
to investigate the origin of the different spin states associated with the two conformations.
First, we optimized the structures of both conformers at U-B3LYP/6-31G(d) level
and calculated their single-point singlet and triplet energies using R(O)-B3LYP/6-31G(d).
The converged energies given in \cref{fig:ni}a and \cref{fig:ni}b are consistent with the spin state ordering observed experimentally.
It is found that in the triplet \textit{cis} conformation the $d_{z^2}$ and $d_{x^2-y^2}$ orbitals of the Ni(II) are singly occupied.
Next, we switch to a simplified model with Ni(II)-porphyrin and a pyridine (\cref{fig:ni}c),
in which the energy dependency of those two orbitals on the  \ch{Ni-N} bond length $d$ (highlighted in \cref{fig:ni}c) was calculated (\cref{fig:orbital}a).
The initial geometry was optimized using U-B3LYP/6-31G(d) with triplet spin state,
which was followed by a rigid scan of the \ch{Ni-N} bond using RO-B3LYP/6-31G(d).
From \cref{fig:orbital}a, it can be seen that the $d_{z^2}$ and $d_{x^2-y^2}$ of the five-coordinate Ni(II) are nearly degenerate,
embracing single occupation with high spin, due to Hund's rules.
As the pyridine departs from the Ni(II), their energies decrease simultaneously but to different extents,
with $d_{z^2}$ much lower than $d_{x^2-y^2}$,
lifting the orbital degeneracy and rendering double occupation of the less energetic $d_{z^2}$.
This phenomenon is consistent with the viewpoint of crystal field theory,
which states that although the fifth ligand raises the energies of all the $d$ orbitals of the center metallic atom due to electrostatic repulsion,
the ligand enhances the head-to-head overlap with the $d_{z^2}$ orbital resulting in the most dramatically elevated energy level near $d_{x^2-y^2}$ which has the highest energy in the presence of the other four ligands (\cref{fig:orbital}b).

\section{6 \ \ Conclusion}

The RO-SCF wavefunction is a useful tool to represent molecular electronic structures with.
However, optimization of RO-SCF orbitals (or equivalently, density) often encounters convergence problems,
which hinders the application of RO wavefunctions to low-spin molecular states.
In this work, we have extended the ARH method to RO-SCF,
which facilitates and enhances its convergence.
The ARH-based RO-SCF method exhibits substantially enhanced optimization performance relative to other algorithms. This improvement arises from the fact that the RO-SCF energy is (approximately) a quadratic function of the density matrices associated with all spin components, and that ARH is capable of approximating the constant Euclidean Hessian and subsequently deriving the corresponding Riemannian Hessian from it in an  exact manner, with only a minimal level of approximation.


For a singlet state with exactly one unpaired $\alpha$ electron and one unpaired $\beta$ electron,
the spin contamination can be resolved by two-determinant (2D) RO-DFT.
2D-DFT considers the equal contributions two broken-symmetry configurations of opposite spin directions of the singly occupied orbitals,
and calculates the energy of the spin pure singlet state using the linear combination of the contaminated state energy and the corresponding triplet energy.
By explicitly accounting for orbital relaxation associated with electronic excitation, 2D-DFT constitutes a powerful methodology for investigating the electronic structure of excited states.
In addition to the conventional 2D-DFT XC energy as a functional (2D\textsubscript{I}-DFT) of the $\alpha$ and $\beta$ electron densities of the mixed singlet and triplet states treated separately
we have developed and implemented an alternative 2D-DFT energy functional, denoted 2D\textsubscript{II}-DFT,
which is explicitly constructed to depend on the total electron density arising from the underlying two-determinant wavefunction.
Our assessment on the ARH performance of 2D\textsubscript{I}- and 2D\textsubscript{II}-DFT has further validated the robustness of the ARH method.
Moreover, the 2D-DFT results surpass both TD-DFT and EOM-CCSD in the accuracy of computed singlet excitation energies for the investigated series of photoactive compounds.

The ARH-based RO-DFT implementation was applied to discuss the spin-crossover process with respect to the variation of singly occupied orbitals.
The calculation reveals that
the azopyridine-substituted Ni-porphyrin (APSNP) exhibits a triplet ground state when the pyridinyl structure coordinates to the nickel(II) center
and a closed-shell singlet ground state in the absence of such pyridinyl-Ni(II) coordination.
The RO-DFT results indicate an increase in the energy of the Ni(II) \(d_{z^2}\) orbital such
that it becomes nearly degenerate with the highest-energy \(d_{x^2-y^2}\) orbital 
when the pyridinyl group approaches the Ni(II) center.
This near-degeneracy leads to a high-spin electronic configuration involving these two orbitals, in accordance with Hund’s rule.

An important contribution of our work is the unified formulation to allow implementation of the direct SCF energy minimization for all spin types, including spin-restricted closed-shell, spin-unrestricted open-shell and spin-restricted open-shell.
To this end, we adopted an extra spin component index to label all the wavefunction variables (density matrices, molecular orbitals, etc).
This new formulation facilitates our ARH-based optimization on the flag manifold.

We anticipate a few directions of our future research topics based on this work.
First, the application of the ARH algorithm can be extended to other SCF wavefunctions confronting convergence difficulties,
such as relativistic DFT.
Second, based on the universal formulation of R/U/RO-SCF,
we will implement the energy derivatives of RO-SCF
as well as RO coupled-perturbed SCF from which the linear response of the density to external perturbations is derived.
Finally, the orthogonality-constrained 2D-DFT of both XC energy types can be implemented
by constraining the overlaps between the targeted state and the lower-lying ones in addition to the flag manifold optimization.\cite{orthogonal}

\section{Code availability}
The algorithm is implemented in \texttt{Chinium},
a quantum chemistry package we are developing,
available at \url{https://github.com/FreemanTheMaverick/Chinium.git}.
It depends on \texttt{Maniverse}, another library of ours, for optimization on manifolds,
which can be found at \url{https://github.com/FreemanTheMaverick/Maniverse.git}.

\section{Supplementary material}
The supplementary material contains the molecular coordinates used in the numerical examples.\edit

\begin{acknowledgement}
We acknowledge Prof.~Wang Zikuan for the discussion on spin-unrestricted DIIS that has coincidentally led to this work.
We also thank Prof.~Tim Kowalczyk for the discussion on the two-determinant restricted open-shell\edit density-functional theory.
The authors acknowledge financial supports from the Hong Kong Research Grants Council through General Research Funds (17307323)
and from the Laboratory for Synthetic Chemistry and Chemical Biology under the Health@InnoHK Program launched by the Innovation and Technology Commission, the Government of HKSAR.
\end{acknowledgement}

\bibliography{reference}

\end{document}


\renewcommand{\thepage}{S\arabic{page}}

\maketitle

\section{1 \ \ Iron(III)-sulfur clusters}

\subsection{1.1 \ \ \textit{c}-\ch{Fe2^{2-}}}

\subsubsection{1.1.1 \ \ Terminal group \ch{-Cl}}
\[\begin{array}{rrrr}
\text{Fe}&                -1.553930&    0.000000&    0.000000\\[-8pt]
 \text{S}&                 0.000000&    0.000000&    1.698662\\[-8pt]
 \text{S}&                 0.000000&    0.000000&   -1.698662\\[-8pt]
\text{Fe}&                 1.553930&    0.000000&    0.000000\\[-8pt]
\text{Cl}&                -2.945325&    1.828799&    0.000000\\[-8pt]
\text{Cl}&                -2.945325&   -1.828799&    0.000000\\[-8pt]
\text{Cl}&                 2.945325&    1.828799&    0.000000\\[-8pt]
\text{Cl}&                 2.945325&   -1.828799&    0.000000
\end{array}\]

\subsubsection{1.1.2 \ \ Terminal group \ch{-SH}}
\[\begin{array}{rrrr}
\text{Fe}& -1.5405202811& 0.& 0.\\[-8pt]
 \text{S}& 0.& 0.& 1.7195381404\\[-8pt]
 \text{S}& 0.& 0.& -1.7195381404\\[-8pt]
\text{Fe}& 1.5405202811& 0.& 0.\\[-8pt]
 \text{S}& 2.9920526606& 1.8694349518& 0.2373229208\\[-8pt]
 \text{S}& -2.9920526606& 1.8694349518& -0.2373229208\\[-8pt]
 \text{S}& -2.9920526606& -1.8694349518& 0.2373229208\\[-8pt]
 \text{S}& 2.9920526606& -1.8694349518& -0.2373229208\\[-8pt]
 \text{H}& -3.8440935550& 1.5312482197& 0.7593879129\\[-8pt]
 \text{H}& 3.8440935550& 1.5312482197& -0.7593879129\\[-8pt]
 \text{H}& -3.8440935550& -1.5312482197& -0.7593879129\\[-8pt]
 \text{H}& 3.8440935550& -1.5312482197& 0.7593879129
\end{array}\]

\subsection{1.2 \ \ \textit{c}-\ch{Fe3^{3-}}}

\subsubsection{1.2.1 \ \ Terminal group \ch{-Cl}}
\[\begin{array}{rrrr}
\text{Fe}&                 0.000000&    0.000000&    3.158673\\[-8pt]
 \text{S}&                 0.000000&    1.697725&    1.638686\\[-8pt]
 \text{S}&                 0.000000&   -1.697725&    1.638686\\[-8pt]
\text{Fe}&                 0.000000&    0.000000&    0.000000\\[-8pt]
\text{Cl}&                 1.829744&    0.000000&    4.632084\\[-8pt]
\text{Cl}&                -1.829744&    0.000000&    4.632084\\[-8pt]
 \text{S}&                 1.697725&    0.000000&   -1.638686\\[-8pt]
 \text{S}&                -1.697725&    0.000000&   -1.638686\\[-8pt]
\text{Fe}&                 0.000000&    0.000000&   -3.158673\\[-8pt]
\text{Cl}&                 0.000000&    1.829744&   -4.632084\\[-8pt]
\text{Cl}&                 0.000000&   -1.829744&   -4.632084
\end{array}\]

\subsubsection{1.2.2 \ \ Terminal group \ch{-SH}}
\[\begin{array}{rrrr}
\text{Fe}& 0.& 0.& 3.1471015812\\[-8pt]
 \text{S}& 1.2021712814& 1.2131110138& 1.6334672953\\[-8pt]
 \text{S}& -1.2021712814& -1.2131110138& 1.6334672953\\[-8pt]
\text{Fe}& 0.& 0.& 0.\\[-8pt]
 \text{S}& 1.2021712814& -1.2131110138& -1.6334672953\\[-8pt]
 \text{S}& -1.2021712814& 1.2131110138& -1.6334672953\\[-8pt]
\text{Fe}& 0.& 0.& -3.1471015812\\[-8pt]
 \text{S}& 1.1628166572& -1.4843083693& 4.6714375358\\[-8pt]
 \text{S}& -1.1628166572& 1.4843083693& 4.6714375358\\[-8pt]
 \text{S}& 1.1628166572& 1.4843083693& -4.6714375358\\[-8pt]
 \text{S}& -1.1628166572& -1.4843083693& -4.6714375358\\[-8pt]
 \text{H}& -1.6526111856& 0.5141593733& 5.4805664100\\[-8pt]
 \text{H}& 1.6526111856& -0.5141593733& 5.4805664100\\[-8pt]
 \text{H}& 1.6526111856& 0.5141593733& -5.4805664100\\[-8pt]
 \text{H}& -1.6526111856& -0.5141593733& -5.4805664100
\end{array}\]

\subsection{1.3 \ \ \textit{c}-\ch{Fe4^{4-}}}

\subsubsection{1.3.1 \ \ Terminal group \ch{-Cl}}
\[\begin{array}{rrrr}
\text{Fe}&                -1.606423&    0.000000&    0.000000\\[-8pt]
\text{S}&                  0.000000&    1.686745&    0.000000\\[-8pt]
\text{S}&                  0.000000&   -1.686745&    0.000000\\[-8pt]
\text{Fe}&                 1.606423&    0.000000&    0.000000\\[-8pt]
\text{S}&                  3.316139&    0.000000&    1.697196\\[-8pt]
\text{S}&                  3.316139&    0.000000&   -1.697196\\[-8pt]
\text{Fe}&                 4.814069&    0.000000&    0.000000\\[-8pt]
\text{Cl}&                 6.353876&    1.829897&    0.000000\\[-8pt]
\text{Cl}&                 6.353876&   -1.829897&    0.000000\\[-8pt]
\text{Fe}&                -4.814069&    0.000000&    0.000000\\[-8pt]
\text{S}&                 -3.316139&    0.000000&   -1.697196\\[-8pt]
\text{S}&                 -3.316139&    0.000000&    1.697196\\[-8pt]
\text{Cl}&                -6.353876&    1.829897&    0.000000\\[-8pt]
\text{Cl}&                -6.353876&   -1.829897&    0.000000
\end{array}\]

\subsubsection{1.3.2 \ \ Terminal group \ch{-SH}}
\[\begin{array}{rrrr}
\text{Fe}& 0.& 0.& -1.6085362346\\[-8pt]
 \text{S}& 0.& 1.6861831898& 0.\\[-8pt]
 \text{S}& 0.& -1.6861831898& 0.\\[-8pt]
\text{Fe}& 0.& 0.& 1.6085362346\\[-8pt]
 \text{S}& -1.7071164044& -0.0063748812& 3.3128046740\\[-8pt]
 \text{S}& 1.7071164044& 0.0063748812& 3.3128046740\\[-8pt]
\text{Fe}& 0.& 0.& 4.8051134459\\[-8pt]
\text{Fe}& 0.& 0.& -4.8051134459\\[-8pt]
 \text{S}& 1.7071164044& -0.0063748812& -3.3128046740\\[-8pt]
 \text{S}& -1.7071164044& 0.0063748812& -3.3128046740\\[-8pt]
 \text{S}& -0.2265752584& 1.8719931195& -6.3881659555\\[-8pt]
 \text{S}& 0.2265752584& 1.8719931195& 6.3881659555\\[-8pt]
 \text{S}& -0.2265752584& -1.8719931195& 6.3881659555\\[-8pt]
 \text{S}& 0.2265752584& -1.8719931195& -6.3881659555\\[-8pt]
 \text{H}& 0.8219025424& 1.525875046& -7.1744623408\\[-8pt]
 \text{H}& -0.8219025424& 1.525875046& 7.1744623408\\[-8pt]
 \text{H}& 0.8219025424& -1.525875046& 7.1744623408\\[-8pt]
 \text{H}& -0.8219025424& -1.525875046& -7.1744623408
\end{array}\]

\subsection{1.4 \ \ \textit{c}-\ch{Fe5^{5-}}}

\subsubsection{1.4.1 \ \ Terminal group \ch{-Cl}}
\[\begin{array}{rrrr}
\text{Fe}& 0.& 0.& 0.\\[-8pt]
\text{S}& 0.& 1.6799746402& -1.6782727266\\[-8pt]
\text{S}& 0.& -1.6799746402& -1.6782727266\\[-8pt]
\text{Fe}& 0.& 0.& -3.2629964097\\[-8pt]
\text{S}& 1.6973711577& 0.& -5.0315105415\\[-8pt]
\text{S}& -1.6973711577& 0.& -5.0315105415\\[-8pt]
\text{Fe}& 0.& 0.& -6.5138886087\\[-8pt]
\text{Cl}& 0.& 1.8326419756& -8.1053223286\\[-8pt]
\text{Cl}& 0.& -1.8326419756& -8.1053223286\\[-8pt]
\text{Fe}& 0.& 0.& 3.2629964097\\[-8pt]
\text{S}& -1.6799746402& 0.& 1.6782727266\\[-8pt]
\text{S}& 1.6799746402& 0.& 1.6782727266\\[-8pt]
\text{Fe}& 0.& 0.& 6.5138886087\\[-8pt]
\text{S}& 0.& -1.6973711577& 5.0315105415\\[-8pt]
\text{S}& 0.& 1.6973711577& 5.0315105415\\[-8pt]
\text{Cl}& -1.8326419756& 0.& 8.1053223286\\[-8pt]
\text{Cl}& 1.8326419756& 0.& 8.1053223286
\end{array}\]

\subsubsection{1.4.2 \ \ Terminal group \ch{-SH}}
\[\begin{array}{rrrr}
\text{Fe}& 0.& 0.& 0.\\[-8pt]
\text{S}& -1.1878990212& -1.1878687763& 1.6785814362\\[-8pt]
\text{S}& 1.1878990212& 1.1878687763& 1.6785814362\\[-8pt]
\text{Fe}& 0.& 0.& 3.2648902418\\[-8pt]
\text{S}& 1.2123955500& -1.2012486326& 5.0286704347\\[-8pt]
\text{S}& -1.2123955500& 1.2012486326& 5.0286704347\\[-8pt]
\text{Fe}& 0.& 0.& 6.5061649742\\[-8pt]
\text{Fe}& 0.& 0.& -3.2648902418\\[-8pt]
\text{S}& -1.1878990212& 1.1878687763& -1.6785814362\\[-8pt]
\text{S}& 1.1878990212& -1.1878687763& -1.6785814362\\[-8pt]
\text{Fe}& 0.& 0.& -6.5061649742\\[-8pt]
\text{S}& 1.2123955500& 1.2012486326& -5.0286704347\\[-8pt]
\text{S}& -1.2123955500& -1.2012486326& -5.0286704347\\[-8pt]
\text{S}& 1.4811484329& -1.1675423956& -8.1395261921\\[-8pt]
\text{S}& -1.4811484329& -1.1675423956& 8.1395261921\\[-8pt]
\text{S}& 1.4811484329& 1.1675423956& 8.1395261921\\[-8pt]
\text{S}& -1.4811484329& 1.1675423956& -8.1395261921\\[-8pt]
\text{H}& -0.4807746691& 1.6573102672& -8.9128083598\\[-8pt]
\text{H}& 0.4807746691& -1.6573102672& -8.9128083598\\[-8pt]
\text{H}& -0.4807746691& -1.6573102672& 8.9128083598\\[-8pt]
\text{H}& 0.4807746691& 1.6573102672& 8.9128083598
\end{array}\]

\subsection{1.5 \ \ \textit{b}-\ch{Fe3^{2-}}}

\subsubsection{1.5.1 \ \ Terminal group \ch{-Cl}}
\[\begin{array}{rrrr}
\text{Fe}&                 0.000000&    1.815368&   -0.090176\\[-8pt]
\text{Fe}&                -1.572155&   -0.907684&   -0.090176\\[-8pt]
\text{Fe}&                 1.572155&   -0.907684&   -0.090176\\[-8pt]
\text{S}&                  0.000000&   -2.261673&   -1.076799\\[-8pt]
\text{S}&                  1.958666&    1.130836&   -1.076799\\[-8pt]
\text{S}&                  0.000000&    0.000000&    1.514807\\[-8pt]
\text{S}&                 -1.958666&    1.130836&   -1.076799\\[-8pt]
\text{Cl}&                 3.430412&   -1.980549&    0.676141\\[-8pt]
\text{Cl}&                -3.430412&   -1.980549&    0.676141\\[-8pt]
\text{Cl}&                 0.000000&    3.961098&    0.676141
\end{array}\]

\subsubsection{1.5.2 \ \ Terminal group \ch{-SH}}
\[\begin{array}{rrrr}
\text{Fe}& -1.5640029405& -0.9029775183& -0.0172434965\\[-8pt]
\text{Fe}& 1.5640029394& -0.9029775201& -0.0172434965\\[-8pt]
\text{Fe}& 0.0000000010& 1.8059550360& -0.0172434965\\[-8pt]
\text{S}& 1.9533800738& 1.1277845091& -1.0326644094\\[-8pt]
\text{S}& -1.9533800725& 1.1277845114& -1.0326644094\\[-8pt]
\text{S}& 0.& -0.0000000008& 1.5981144850\\[-8pt]
\text{S}& -0.0000000013& -2.2555690230& -1.0326644094\\[-8pt]
\text{S}& -3.4941859782& -2.0173692133& 0.7568660972\\[-8pt]
\text{S}& 0.0000000023& 4.0347384280& 0.7568660972\\[-8pt]
\text{S}& 3.4941859759& -2.0173692172& 0.7568660972\\[-8pt]
\text{H}& 0.0000000026& 4.5901951309& -0.4781770100\\[-8pt]
\text{H}& -3.9752255937& -2.2950975644& -0.4781770100\\[-8pt]
\text{H}& 3.9752255911& -2.2950975690& -0.4781770100
\end{array}\]

\subsection{1.6 \ \ \textit{b}-\ch{Fe4}}

\subsubsection{1.6.1 \ \ Terminal group \ch{-Cl}}
\[\begin{array}{rrrr}
\text{Fe}&                 1.147932&    1.147932&    1.147932\\[-8pt]
\text{Fe}&                -1.147932&   -1.147932&    1.147932\\[-8pt]
\text{Fe}&                -1.147932&    1.147932&   -1.147932\\[-8pt]
\text{Fe}&                 1.147932&   -1.147932&   -1.147932\\[-8pt]
\text{S}&                 -1.241254&   -1.241254&   -1.241254\\[-8pt]
\text{S}&                  1.241254&    1.241254&   -1.241254\\[-8pt]
\text{S}&                  1.241254&   -1.241254&    1.241254\\[-8pt]
\text{S}&                 -1.241254&    1.241254&    1.241254\\[-8pt]
\text{Cl}&                 2.385819&   -2.385819&   -2.385819\\[-8pt]
\text{Cl}&                -2.385819&   -2.385819&    2.385819\\[-8pt]
\text{Cl}&                 2.385819&    2.385819&    2.385819\\[-8pt]
\text{Cl}&                -2.385819&    2.385819&   -2.385819
\end{array}\]

\subsubsection{1.6.2 \ \ Terminal group \ch{-SH}}
\[\begin{array}{rrrr}
\text{Fe}& 0.1174068478& -0.0245669830& 2.0319311998\\[-8pt]
\text{Fe}& -0.0766739114& -1.8002291156& -0.6547616412\\[-8pt]
\text{Fe}& -1.5436138341& 1.0697580253& -0.5153876300\\[-8pt]
\text{Fe}& 1.6760522610& 0.9094134382& -0.6353457981\\[-8pt]
\text{S}& -0.0180073773& 0.1388452860& -2.1069288756\\[-8pt]
\text{S}& 0.1824740724& 2.0482572111& 0.8652865114\\[-8pt]
\text{S}& 1.7657889872& -1.1052831393& 0.6843637429\\[-8pt]
\text{S}& -1.7643378565& -0.9052223288& 0.7824677013\\[-8pt]
\text{S}& 3.5006577157& 1.9090038346& -1.4391504160\\[-8pt]
\text{S}& -0.2059657839& -3.8812994879& -1.4428034154\\[-8pt]
\text{S}& 0.2153851998& -0.0863530933& 4.2588030160\\[-8pt]
\text{S}& -3.3491573156& 2.2147880129& -1.1525081989\\[-8pt]
\text{H}& -1.5342741617& -3.9158732451& -1.6992125659\\[-8pt]
\text{H}& 4.0774203499& 2.2567674884& -0.2654101648\\[-8pt]
\text{H}& -0.2826129940& 1.1462404477& 4.5117501528\\[-8pt]
\text{H}& -2.7605421991& 3.0414313688& -2.0477027382
\end{array}\]

\section{2 \ \ Photo-active compounds}

\subsection{2.1 \ \ Alloxazine}
\[\begin{array}{rrrr}
\text{C}&-3.7937665109&-0.6933001606&0.\\[-8pt]
\text{C}&-2.6119263077&-1.3879983298&0.\\[-8pt]
\text{C}&-1.3856541600&-0.6877218888&0.\\[-8pt]
\text{C}&-1.4017996875&0.7417790638&0.\\[-8pt]
\text{C}&-2.6388482057&1.4281647880&0.\\[-8pt]
\text{C}&-3.8109078790&0.7211479782&0.\\[-8pt]
\text{C}&0.8739095828&0.7735410201&0.\\[-8pt]
\text{C}&0.8822739509&-0.6546628452&0.\\[-8pt]
\text{C}&3.3311566699&-0.7109165116&0.\\[-8pt]
\text{C}&2.1686810060&1.5159093466&0.\\[-8pt]
\text{H}&-2.5846891681&-2.4684709750&0.\\[-8pt]
\text{H}&-2.6172545383&2.5087959010&0.\\[-8pt]
\text{N}&-0.2476987574&1.4456840597&0.\\[-8pt]
\text{N}&-0.2145559667&-1.3696847881&0.\\[-8pt]
\text{N}&2.0913669019&-1.3168588855&0.\\[-8pt]
\text{N}&3.2840344364&0.6776542709&0.\\[-8pt]
\text{O}&4.3693327155&-1.3309785095&0.\\[-8pt]
\text{O}&2.2848881380&2.7160197980&0.\\[-8pt]
\text{H}&-4.7317998527&-1.2321923139&0.\\[-8pt]
\text{H}&-4.7590311633&1.2408750137&0.\\[-8pt]
\text{H}&2.0815621561&-2.3257991475&0.\\[-8pt]
\text{H}&4.1849223197&1.1349584355&0.
\end{array}\]

\subsection{2.2 \ \ Isoalloxazine}
\[\begin{array}{rrrr}
\text{C}&-4.0770504327&-0.7625141725&0.\\[-8pt]
\text{C}&-2.8960522845&-1.4796248677&0.\\[-8pt]
\text{C}&-1.6809660601&-0.7903883665&0.\\[-8pt]
\text{C}&-1.6616710975&0.6238626846&0.\\[-8pt]
\text{C}&-2.8782885050&1.3240244366&0.\\[-8pt]
\text{C}&-4.0739897001&0.6393313743&0.\\[-8pt]
\text{C}&0.6369620314&0.6704945848&0.\\[-8pt]
\text{C}&0.7334809454&-0.7868483913&0.\\[-8pt]
\text{C}&3.0374068876&-0.8394125124&0.\\[-8pt]
\text{C}&1.9347722973&1.4197927078&0.\\[-8pt]
\text{H}&-5.0181547032&-1.2952204989&0.\\[-8pt]
\text{H}&-2.9043924928&-2.5618178862&0.\\[-8pt]
\text{H}&-2.8355009335&2.4038068400&0.\\[-8pt]
\text{H}&-5.0101839239&1.1788601014&0.\\[-8pt]
\text{H}&3.9341557333&1.0127428592&0.\\[-8pt]
\text{N}&-0.4793276525&1.3189173015&0.\\[-8pt]
\text{N}&-0.4687420731&-1.4347743481&0.\\[-8pt]
\text{N}&1.8195218032&-1.4938719336&0.\\[-8pt]
\text{N}&3.0229052541&0.5752838539&0.\\[-8pt]
\text{H}&-0.4241979643&-2.4451199472&0.\\[-8pt]
\text{O}&4.1020522940&-1.4122479204&0.\\[-8pt]
\text{O}&2.0290667767&2.6247497908&0.
\end{array}\]

\subsection{2.3 \ \ Benzaldehyde}
\[\begin{array}{rrrr}
\text{C}&-1.2831747934&0.3622651339&0.\\[-8pt]
\text{C}&-0.2105085507&-0.5295497361&0.\\[-8pt]
\text{C}&1.0992090920&-0.0414116172&0.\\[-8pt]
\text{C}&1.3269498513&1.3245567244&0.\\[-8pt]
\text{C}&0.2509128424&2.2112546425&0.\\[-8pt]
\text{C}&-1.0542348360&1.7318329765&0.\\[-8pt]
\text{H}&-2.2959809739&-0.0223220977&0.\\[-8pt]
\text{H}&1.9167159174&-0.7490988503&0.\\[-8pt]
\text{H}&2.3394900580&1.7049571331&0.\\[-8pt]
\text{H}&0.4323570170&3.2778808599&0.\\[-8pt]
\text{H}&-1.8862778574&2.4226686137&0.\\[-8pt]
\text{C}&-0.4726305715&-1.9838830626&0.\\[-8pt]
\text{O}&0.3781409022&-2.8419805513&0.\\[-8pt]
\text{H}&-1.5483806174&-2.2557505587&0.
\end{array}\]

\subsection{2.4 \ \ Benzophenone}
\[\begin{array}{rrrr}
\text{C}&0.&0.&1.0589173156\\[-8pt]
\text{O}&0.&0.&2.2772986932\\[-8pt]
\text{C}&-0.0943515101&1.2963629666&0.3135877542\\[-8pt]
\text{C}&0.5126562415&1.5007359496&-0.9279288654\\[-8pt]
\text{C}&-0.7532926322&2.3631179350&0.9320464417\\[-8pt]
\text{C}&0.4575087451&2.7472592625&-1.5397755622\\[-8pt]
\text{H}&1.0506928720&0.6942983814&-1.4053712925\\[-8pt]
\text{C}&-0.8273500721&3.5993987691&0.3102510932\\[-8pt]
\text{H}&-1.1979904010&2.2020405406&1.9038525415\\[-8pt]
\text{C}&-0.2199263561&3.7945263526&-0.9273150514\\[-8pt]
\text{H}&0.9447000699&2.9001254948&-2.4934375818\\[-8pt]
\text{H}&-1.3517614434&4.4145268317&0.7906815722\\[-8pt]
\text{H}&-0.2704307489&4.7619272982&-1.4091997544\\[-8pt]
\text{C}&0.0943515101&-1.2963629666&0.3135877542\\[-8pt]
\text{C}&-0.5126562415&-1.5007359496&-0.9279288654\\[-8pt]
\text{C}&0.7532926322&-2.3631179350&0.9320464417\\[-8pt]
\text{C}&-0.4575087451&-2.7472592625&-1.5397755622\\[-8pt]
\text{H}&-1.0506928720&-0.6942983814&-1.4053712925\\[-8pt]
\text{C}&0.8273500721&-3.5993987691&0.3102510932\\[-8pt]
\text{H}&1.1979904010&-2.2020405406&1.9038525415\\[-8pt]
\text{C}&0.2199263561&-3.7945263526&-0.9273150514\\[-8pt]
\text{H}&-0.9447000699&-2.9001254948&-2.4934375818\\[-8pt]
\text{H}&1.3517614434&-4.4145268317&0.7906815722\\[-8pt]
\text{H}&0.2704307489&-4.7619272982&-1.4091997544
\end{array}\]

\subsection{2.5 \ \ Xanthone}
\[\begin{array}{rrrr}
\text{C}&0.&3.5699851734&-0.9373052323\\[-8pt]
\text{C}&0.&2.3424512420&-1.5716646609\\[-8pt]
\text{C}&0.&1.1803004781&-0.8009910981\\[-8pt]
\text{C}&0.&1.2390483000&0.5954541018\\[-8pt]
\text{C}&0.&2.4969043045&1.2127641121\\[-8pt]
\text{C}&0.&3.6526030058&0.4597701033\\[-8pt]
\text{C}&0.&0.&1.3950623076\\[-8pt]
\text{C}&0.&-1.2390483000&0.5954541018\\[-8pt]
\text{C}&0.&-1.1803004781&-0.8009910981\\[-8pt]
\text{C}&0.&-2.3424512420&-1.5716646609\\[-8pt]
\text{H}&0.&-2.2585730539&-2.6490570218\\[-8pt]
\text{C}&0.&-3.5699851734&-0.9373052323\\[-8pt]
\text{C}&0.&-3.6526030058&0.4597701033\\[-8pt]
\text{C}&0.&-2.4969043045&1.2127641121\\[-8pt]
\text{H}&0.&4.4741336308&-1.5311870513\\[-8pt]
\text{H}&0.&2.2585730539&-2.6490570218\\[-8pt]
\text{H}&0.&2.5224307605&2.2934181666\\[-8pt]
\text{H}&0.&4.6191429523&0.9439452532\\[-8pt]
\text{H}&0.&-4.4741336308&-1.5311870513\\[-8pt]
\text{H}&0.&-4.6191429523&0.9439452532\\[-8pt]
\text{H}&0.&-2.5224307605&2.2934181666\\[-8pt]
\text{O}&0.&0.&2.6166037281\\[-8pt]
\text{O}&0.&0.&-1.4854449209
\end{array}\]

\subsection{2.6 \ \ Thioxanthone}
\[\begin{array}{rrrr}
\text{C}&0.&3.7879544354&-0.6849326415\\[-8pt]
\text{C}&0.&2.6382099988&-1.4490469901\\[-8pt]
\text{C}&0.&1.3803499366&-0.8309721243\\[-8pt]
\text{C}&0.&1.2825366928&0.5671304316\\[-8pt]
\text{C}&0.&2.4688303369&1.3185246166\\[-8pt]
\text{C}&0.&3.7058099842&0.7099054492\\[-8pt]
\text{C}&0.&0.&1.3141775860\\[-8pt]
\text{C}&0.&-1.2825366928&0.5671304316\\[-8pt]
\text{C}&0.&-1.3803499366&-0.8309721243\\[-8pt]
\text{C}&0.&-2.6382099988&-1.4490469901\\[-8pt]
\text{H}&0.&-2.7026436383&-2.5291851010\\[-8pt]
\text{C}&0.&-3.7879544354&-0.6849326415\\[-8pt]
\text{C}&0.&-3.7058099842&0.7099054492\\[-8pt]
\text{C}&0.&-2.4688303369&1.3185246166\\[-8pt]
\text{H}&0.&4.7529095693&-1.1740466405\\[-8pt]
\text{H}&0.&2.7026436383&-2.5291851010\\[-8pt]
\text{H}&0.&2.3706367256&2.3943381967\\[-8pt]
\text{H}&0.&4.6066382173&1.3078369723\\[-8pt]
\text{H}&0.&-4.7529095693&-1.1740466405\\[-8pt]
\text{H}&0.&-4.6066382173&1.3078369723\\[-8pt]
\text{H}&0.&-2.3706367256&2.3943381967\\[-8pt]
\text{S}&0.&0.&-1.9134193658\\[-8pt]
\text{O}&0.&0.&2.5372966818
\end{array}\]

\subsection{2.7 \ \ Acridone}
\[\begin{array}{rrrr}
\text{C}&0.&3.6189909672&-0.9052856826\\[-8pt]
\text{C}&0.&2.4102416525&-1.5690210086\\[-8pt]
\text{C}&0.&1.2114102405&-0.8386527019\\[-8pt]
\text{C}&0.&1.2478296833&0.5676404787\\[-8pt]
\text{C}&0.&2.4908896353&1.2137023740\\[-8pt]
\text{C}&0.&3.6672007144&0.4945021627\\[-8pt]
\text{C}&0.&0.&1.3552698555\\[-8pt]
\text{C}&0.&-1.2478296833&0.5676404787\\[-8pt]
\text{C}&0.&-1.2114102405&-0.8386527019\\[-8pt]
\text{C}&0.&-2.4102416525&-1.5690210086\\[-8pt]
\text{H}&0.&-2.3782360672&-2.6516603678\\[-8pt]
\text{C}&0.&-3.6189909672&-0.9052856826\\[-8pt]
\text{C}&0.&-3.6672007144&0.4945021627\\[-8pt]
\text{C}&0.&-2.4908896353&1.2137023740\\[-8pt]
\text{H}&0.&4.5377473343&-1.4766790672\\[-8pt]
\text{H}&0.&2.3782360672&-2.6516603678\\[-8pt]
\text{H}&0.&2.4871402022&2.2946709333\\[-8pt]
\text{H}&0.&4.6208638501&1.0034013473\\[-8pt]
\text{H}&0.&-4.5377473343&-1.4766790672\\[-8pt]
\text{H}&0.&-4.6208638501&1.0034013473\\[-8pt]
\text{H}&0.&-2.4871402022&2.2946709333\\[-8pt]
\text{O}&0.&0.&2.5805266581\\[-8pt]
\text{N}&0.&0.&-1.4909906716\\[-8pt]
\text{H}&0.&0.&-2.4968676774
\end{array}\]

\section{3 \ \ Azopyridine-substituted Ni-porphyrin}

\subsection{3.1 \ \ \textit{trans}-APSNP}
\[\begin{array}{rrrr}
\text{N}&                -1.410965&  -0.113153&   1.473818\\[-8pt]
\text{N}&                -2.893860&   1.217050&  -0.375966\\[-8pt]
\text{N}&                -4.309926&  -1.085353&  -0.901421\\[-8pt]
\text{N}&                -2.818198&  -2.418713&   0.960883\\[-8pt]
\text{C}&                -0.940087&  -0.838747&   2.548874\\[-8pt]
\text{C}&                 0.124836&  -0.141518&   3.215836\\[-8pt]
\text{C}&                 0.341801&   1.001014&   2.514450\\[-8pt]
\text{C}&                -0.628978&   1.025359&   1.448558\\[-8pt]
\text{C}&                -0.813949&   2.115249&   0.599829\\[-8pt]
\text{C}&                -1.942848&   2.207891&  -0.213515\\[-8pt]
\text{C}&                -2.326168&   3.401615&  -0.926600\\[-8pt]
\text{C}&                -3.544123&   3.154764&  -1.477347\\[-8pt]
\text{C}&                -3.873817&   1.794244&  -1.157348\\[-8pt]
\text{C}&                -4.966703&   1.133756&  -1.689292\\[-8pt]
\text{C}&                -5.132180&  -0.235161&  -1.610418\\[-8pt]
\text{C}&                -6.109478&  -0.971922&  -2.368021\\[-8pt]
\text{C}&                -5.854240&  -2.289575&  -2.152592\\[-8pt]
\text{C}&                -4.756346&  -2.350062&  -1.223353\\[-8pt]
\text{C}&                -4.300546&  -3.520585&  -0.645887\\[-8pt]
\text{C}&                -3.426834&  -3.534607&   0.425235\\[-8pt]
\text{C}&                -3.141110&  -4.706370&   1.211305\\[-8pt]
\text{C}&                -2.386646&  -4.293775&   2.264102\\[-8pt]
\text{C}&                -2.171417&  -2.881737&   2.086806\\[-8pt]
\text{C}&                -1.332166&  -2.122747&   2.879395\\[-8pt]
\text{H}&                 0.646100&  -0.508962&   4.090442\\[-8pt]
\text{H}&                 1.065618&   1.779963&   2.704576\\[-8pt]
\text{H}&                -1.742649&   4.309692&  -0.966994\\[-8pt]
\text{H}&                -4.160982&   3.811277&  -2.077578\\[-8pt]
\text{H}&                -5.670383&   1.705182&  -2.285718\\[-8pt]
\text{H}&                -6.865414&  -0.520265&  -2.997598\\[-8pt]
\text{H}&                -6.366392&  -3.155061&  -2.553174\\[-8pt]
\text{H}&                -4.724717&  -4.461691&  -0.980048\\[-8pt]
\text{H}&                -3.505128&  -5.700289&   0.984541\\[-8pt]
\text{H}&                -1.981665&  -4.878156&   3.080331\\[-8pt]
\text{H}&                -0.878489&  -2.590330&   3.747066\\[-8pt]
\text{Ni}&               -2.856025&  -0.600530&   0.287460\\[-8pt]
\text{C}&                 0.152312&   3.259994&   0.666474\\[-8pt]
\text{C}&                -0.224861&   4.422329&   1.357420
\end{array}\]
\[\begin{array}{rrrr}
\text{C}&                 1.448883&   3.185784&   0.103073\\[-8pt]
\text{C}&                 0.646326&   5.501315&   1.499725\\[-8pt]
\text{H}&                -1.215753&   4.463893&   1.800917\\[-8pt]
\text{C}&                 2.316069&   4.280756&   0.262321\\[-8pt]
\text{C}&                 1.926271&   5.427868&   0.949909\\[-8pt]
\text{H}&                 0.328482&   6.387697&   2.041839\\[-8pt]
\text{H}&                 3.305828&   4.230336&  -0.182928\\[-8pt]
\text{H}&                 2.617361&   6.260399&   1.049909\\[-8pt]
\text{C}&                 1.940401&   1.994574&  -0.648290\\[-8pt]
\text{C}&                 3.183456&   1.431334&  -0.329574\\[-8pt]
\text{C}&                 1.213224&   1.431527&  -1.709551\\[-8pt]
\text{C}&                 3.686294&   0.333212&  -1.038430\\[-8pt]
\text{H}&                 3.779173&   1.824658&   0.488463\\[-8pt]
\text{C}&                 1.713414&   0.334342&  -2.420003\\[-8pt]
\text{H}&                 0.257875&   1.860676&  -1.992264\\[-8pt]
\text{C}&                 2.943698&  -0.222943&  -2.093531\\[-8pt]
\text{H}&                 1.131427&  -0.080246&  -3.238604\\[-8pt]
\text{H}&                 3.345529&  -1.074088&  -2.631159\\[-8pt]
\text{N}&                 4.949797&  -0.135876&  -0.602161\\[-8pt]
\text{N}&                 5.400192&  -1.123662&  -1.243893\\[-8pt]
\text{C}&                 6.663503&  -1.587850&  -0.801690\\[-8pt]
\text{C}&                 7.405003&  -1.046671&   0.266186\\[-8pt]
\text{C}&                 7.202808&  -2.676798&  -1.494939\\[-8pt]
\text{N}&                 8.592528&  -1.519282&   0.646552\\[-8pt]
\text{H}&                 7.005248&  -0.202030&   0.819749\\[-8pt]
\text{C}&                 8.442161&  -3.175882&  -1.104071\\[-8pt]
\text{H}&                 6.639909&  -3.103719&  -2.319840\\[-8pt]
\text{C}&                 9.093704&  -2.564209&  -0.032093\\[-8pt]
\text{H}&                 8.896322&  -4.019313&  -1.615392\\[-8pt]
\text{H}&                10.064402&  -2.926477&   0.302188\\[-8pt]
\end{array}\]

\subsection{3.2 \ \ \textit{cis}-APSNP}
\[\begin{array}{rrrr}
\text{N}&                 0.179840&  -0.594987&   1.737896\\[-8pt]
\text{N}&                 0.057556&  -1.518763&  -0.952466\\[-8pt]
\text{N}&                 2.927453&  -1.229181&  -1.199211\\[-8pt]
\text{N}&                 3.050136&  -0.292863&   1.504297\\[-8pt]
\text{C}&                 0.451040&  -0.163607&   3.014061\\[-8pt]
\text{C}&                -0.762684&  -0.116751&   3.793660\\[-8pt]
\text{C}&                -1.773224&  -0.525017&   2.975708\\[-8pt]
\text{C}&                -1.176007&  -0.828714&   1.690414\\[-8pt]
\text{C}&                -1.882452&  -1.290774&   0.565928\\[-8pt]
\text{C}&                -1.278626&  -1.641612&  -0.654811\\[-8pt]
\text{C}&                -1.982369&  -2.230860&  -1.777284\\[-8pt]
\text{C}&                -1.057229&  -2.453015&  -2.749934\\[-8pt]
\text{C}&                 0.212416&  -2.005024&  -2.226987\\[-8pt]
\text{C}&                 1.416790&  -2.093364&  -2.924190\\[-8pt]
\text{C}&                 2.678784&  -1.749452&  -2.443924\\[-8pt]
\text{C}&                 3.916987&  -1.922135&  -3.170700\\[-8pt]
\text{C}&                 4.917721&  -1.505650&  -2.344211\\[-8pt]
\text{C}&                 4.290047&  -1.079735&  -1.112425\\[-8pt]
\text{C}&                 4.970775&  -0.603881&   0.008300\\[-8pt]
\text{C}&                 4.392803&  -0.249730&   1.226931\\[-8pt]
\text{C}&                 5.131026&   0.188939&   2.392308\\[-8pt]
\text{C}&                 4.215033&   0.407468&   3.375958\\[-8pt]
\text{C}&                 2.918166&   0.102011&   2.810467\\[-8pt]
\text{C}&                 1.712494&   0.173260&   3.505223\\[-8pt]
\text{H}&                -0.822307&   0.181240&   4.833145\\[-8pt]
\text{H}&                -2.821909&  -0.630889&   3.216019\\[-8pt]
\text{H}&                -3.037809&  -2.462667&  -1.797395\\[-8pt]
\text{H}&                -1.204855&  -2.894791&  -3.727700\\[-8pt]
\text{H}&                 1.368407&  -2.500020&  -3.930653\\[-8pt]
\text{H}&                 3.997813&  -2.323848&  -4.173240\\[-8pt]
\text{H}&                 5.984296&  -1.497310&  -2.532420\\[-8pt]
\text{H}&                 6.052218&  -0.529299&  -0.065547\\[-8pt]
\text{H}&                 6.207462&   0.298562&   2.439945\\[-8pt]
\text{H}&                 4.387026&   0.733375&   4.394329\\[-8pt]
\text{H}&                 1.760394&   0.503147&   4.539350\\[-8pt]
\text{Ni}&                1.519919&  -0.632083&   0.176580\\[-8pt]
\text{C}&                -3.372356&  -1.456793&   0.676632\\[-8pt]
\text{C}&                -3.895836&  -2.582863&   1.329239\\[-8pt]
\text{C}&                -4.267748&  -0.528896&   0.094609\\[-8pt]
\text{C}&                -5.271307&  -2.803888&   1.404936
\end{array}\]
\[\begin{array}{rrrr}
\text{H}&                -3.205549&  -3.297528&   1.768775\\[-8pt]
\text{C}&                -5.647787&  -0.770144&   0.164727\\[-8pt]
\text{C}&                -6.151393&  -1.896640&   0.814700\\[-8pt]
\text{H}&                -5.650879&  -3.685349&   1.914673\\[-8pt]
\text{H}&                -6.329080&  -0.052281&  -0.284477\\[-8pt]
\text{H}&                -7.224583&  -2.060280&   0.863111\\[-8pt]
\text{C}&                -3.772450&   0.701830&  -0.593843\\[-8pt]
\text{C}&                -3.182380&   1.735256&   0.144112\\[-8pt]
\text{C}&                -3.890945&   0.844118&  -1.985152\\[-8pt]
\text{C}&                -2.638204&   2.843368&  -0.509852\\[-8pt]
\text{H}&                -3.113626&   1.665178&   1.225259\\[-8pt]
\text{C}&                -3.382106&   1.972627&  -2.628202\\[-8pt]
\text{H}&                -4.364156&   0.055951&  -2.562831\\[-8pt]
\text{C}&                -2.735799&   2.971011&  -1.900168\\[-8pt]
\text{H}&                -3.474141&   2.069038&  -3.706525\\[-8pt]
\text{H}&                -2.332755&   3.850079&  -2.393791\\[-8pt]
\text{N}&                -2.118624&   3.935577&   0.273284\\[-8pt]
\text{N}&                -0.909056&   4.239961&   0.265461\\[-8pt]
\text{C}&                 0.095327&   3.426198&  -0.359754\\[-8pt]
\text{C}&                 0.263586&   2.075914&  -0.029979\\[-8pt]
\text{C}&                 1.071773&   4.049500&  -1.141986\\[-8pt]
\text{N}&                 1.307653&   1.353252&  -0.457048\\[-8pt]
\text{H}&                -0.431120&   1.566414&   0.624074\\[-8pt]
\text{C}&                 2.132109&   3.283964&  -1.617127\\[-8pt]
\text{H}&                 0.986831&   5.108937&  -1.365158\\[-8pt]
\text{C}&                 2.218724&   1.942178&  -1.251176\\[-8pt]
\text{H}&                 2.900166&   3.722180&  -2.246049\\[-8pt]
\text{H}&                 3.037529&   1.312945&  -1.579863
\end{array}\]

\subsection{3.3 \ \ The simplified model}
\[\begin{array}{rrrr}
\text{N}&                 1.436201&   1.440851&   0.675834\\[-8pt]
\text{N}&                 1.436201&  -1.440851&   0.675834\\[-8pt]
\text{N}&                -1.436201&  -1.440851&   0.675834\\[-8pt]
\text{N}&                -1.436201&   1.440851&   0.675834\\[-8pt]
\text{C}&                 1.244270&   2.798124&   0.739647\\[-8pt]
\text{C}&                 2.519104&   3.480343&   0.799353\\[-8pt]
\text{C}&                 3.480771&   2.515533&   0.773243\\[-8pt]
\text{C}&                 2.793626&   1.244142&   0.700007\\[-8pt]
\text{C}&                 3.424059&   0.000000&   0.692087\\[-8pt]
\text{C}&                 2.793626&  -1.244142&   0.700007\\[-8pt]
\text{C}&                 3.480771&  -2.515533&   0.773243\\[-8pt]
\text{C}&                 2.519104&  -3.480343&   0.799353\\[-8pt]
\text{C}&                 1.244270&  -2.798124&   0.739647\\[-8pt]
\text{C}&                 0.000000&  -3.430445&   0.757542\\[-8pt]
\text{C}&                -1.244270&  -2.798124&   0.739647\\[-8pt]
\text{C}&                -2.519104&  -3.480343&   0.799353\\[-8pt]
\text{C}&                -3.480771&  -2.515533&   0.773243\\[-8pt]
\text{C}&                -2.793626&  -1.244142&   0.700007\\[-8pt]
\text{C}&                -3.424059&   0.000000&   0.692087\\[-8pt]
\text{C}&                -2.793626&   1.244142&   0.700007\\[-8pt]
\text{C}&                -3.480771&   2.515533&   0.773243\\[-8pt]
\text{C}&                -2.519104&   3.480343&   0.799353\\[-8pt]
\text{C}&                -1.244270&   2.798124&   0.739647\\[-8pt]
\text{C}&                 0.000000&   3.430445&   0.757542\\[-8pt]
\text{H}&                 2.646381&   4.553961&   0.864741\\[-8pt]
\text{H}&                 4.556135&   2.638322&   0.813141\\[-8pt]
\text{H}&                 4.510403&   0.000000&   0.715551\\[-8pt]
\text{H}&                 4.556135&  -2.638322&   0.813141\\[-8pt]
\text{H}&                 2.646381&  -4.553961&   0.864741\\[-8pt]
\text{H}&                 0.000000&  -4.515781&   0.810817\\[-8pt]
\text{H}&                -2.646381&  -4.553961&   0.864741\\[-8pt]
\text{H}&                -4.556135&  -2.638322&   0.813141\\[-8pt]
\text{H}&                -4.510403&   0.000000&   0.715551\\[-8pt]
\text{H}&                -4.556135&   2.638322&   0.813141\\[-8pt]
\text{H}&                -2.646381&   4.553961&   0.864741\\[-8pt]
\text{H}&                 0.000000&   4.515781&   0.810817\\[-8pt]
\text{Ni}&                0.000000&   0.000000&   0.394979\\[-8pt]
\text{N}&                 0.000000&   0.000000&  -1.673720\\[-8pt]
\text{C}&                 0.000000&  -1.154778&  -2.360232\\[-8pt]
\text{C}&                 0.000000&   1.154778&  -2.360232
\end{array}\]
\[\begin{array}{rrrr}
\text{C}&                 0.000000&  -1.199508&  -3.751805\\[-8pt]
\text{H}&                 0.000000&  -2.060625&  -1.764106\\[-8pt]
\text{C}&                 0.000000&   1.199508&  -3.751805\\[-8pt]
\text{H}&                 0.000000&   2.060625&  -1.764106\\[-8pt]
\text{C}&                 0.000000&   0.000000&  -4.462912\\[-8pt]
\text{H}&                 0.000000&  -2.158135&  -4.260511\\[-8pt]
\text{H}&                 0.000000&   2.158135&  -4.260511\\[-8pt]
\text{H}&                 0.000000&   0.000000&  -5.549380
\end{array}\]